\documentclass[
aps,prd,
showpacs,twocolumn,notitlepage,
amssymb,amsmath,amsfonts,mathrsfs,
nofootinbib,superscriptaddress,
floats,floatfix
]{revtex4-2}

\usepackage{graphicx}
\usepackage{xcolor}
\usepackage[colorlinks=true]{hyperref}
\hypersetup{citecolor=cyan,linkcolor=magenta}
\usepackage{multirow,array}

\usepackage{amsmath}
\usepackage{graphics}
\usepackage{epstopdf}
\usepackage{enumerate}

\usepackage{savesym}
\savesymbol{tablenum}
\usepackage{siunitx}
\restoresymbol{SIX}{tablenum}

\usepackage{hyperref}

\newcommand{\mnras}{Mon. Not. R. Astron. Soc.}
\newcommand{\pasj}{Publ. Astron. Soc. Jpn.}
\newcommand{\apjs}{Astrophys. J. Suppl. Ser.}
\newcommand{\apjl}{Astrophys. J. Lett.}
\newcommand{\aap}{Astron. Astrophys.}

\newcommand{\beq}{\begin{equation}}
\newcommand{\eeq}{\end{equation}}
\newcommand{\beqn}{\begin{eqnarray}}
\newcommand{\eeqn}{\end{eqnarray}}

\begin{document}


\title{Self-consistent scenario for jet and stellar explosion in collapsar: \\General relativistic magnetohydrodynamics simulation 
with dynamo}

\author{Masaru Shibata}
\affiliation{Max-Planck-Institut f\"ur Gravitationsphysik (Albert-Einstein-Institut), Am M\"uhlenberg 1, D-14476 Potsdam-Golm, Germany}
\affiliation{Center for Gravitational Physics and Quantum Information, Yukawa Institute for Theoretical Physics, Kyoto University, Kyoto, 606-8502, Japan}

\author{Sho Fujibayashi}
\affiliation{Max-Planck-Institut f\"ur Gravitationsphysik (Albert-Einstein-Institut), Am M\"uhlenberg 1, D-14476 Potsdam-Golm, Germany}
\affiliation{Frontier Research Institute for Interdisciplinary Sciences, Tohoku University, Sendai 980-8578, Japan}
\affiliation{Astronomical Institute, Tohoku University, Aoba, Sendai 980-8578, Japan}

\author{Shinya Wanajo}
\affiliation{Max-Planck-Institut f\"ur Gravitationsphysik (Albert-Einstein-Institut), Am M\"uhlenberg 1, D-14476 Potsdam-Golm, Germany}
\affiliation{Astronomical Institute, Tohoku University, Aoba, Sendai 980-8578, Japan}

\author{Kunihito Ioka}
\affiliation{Center for Gravitational Physics and Quantum Information, Yukawa Institute for Theoretical Physics, Kyoto University, Kyoto, 606-8502, Japan}

\author{Alan Tsz-Lok Lam}
\affiliation{Max-Planck-Institut f\"ur Gravitationsphysik (Albert-Einstein-Institut), Am M\"uhlenberg 1, D-14476 Potsdam-Golm, Germany}

\author{Yuichiro Sekiguchi}
\affiliation{Department of Physics, Toho University, Funabashi, Chiba 274-8510, Japan}

\date{\today}


\begin{abstract}
A resistive magnetohydrodynamics simulation with a dynamo term is performed for modeling the collapsar in full general relativity. As an initial condition, a spinning black hole and infalling stellar matter are modeled based on a stellar evolution result, superimposing a weak toroidal magnetic field. 
After the growth of a massive torus around the black hole, the magnetic field is amplified in it, developing poloidal fields via dynamo. In an early stage of the torus growth, magnetic fluxes that fall to the vicinity of the central black hole are swallowed by the black hole and global poloidal magnetic fields that can be the source of the Blandford-Znajek mechanism are not developed. However, in a later stage in which the ram pressure of the infalling matter becomes weak, the magnetic field amplified by the black hole spin via the winding becomes large enough to expel the infalling matter by the magnetic pressure, and subsequently, a global poloidal magnetic field that penetrates the black hole is established, launching a jet along the spin axis by the Blandford-Znajek mechanism with the luminosity suitable for explaining typical long gamma-ray bursts. Together with the jet launch, the effectively viscous effect in the inner region of the torus and the magnetocentrifugal effect drive the stellar explosion with the explosion energy comparable to typical or powerful supernovae. We also find large amounts of synthesized $^{56}$Ni and Zn associated with the stellar explosion. In the presence of jet launching, $r$-process elements are weakly synthesized. The numerical results of the explosion energy, ejecta mass, and $^{56}$Ni mass are in a good agreement with those for observed broad-lined type Ic supernovae. Our result illustrates a self-consistent scenario for the gamma-ray-burst-associated broad-lined type Ic supernovae. 
\end{abstract} 


\maketitle

\section{Introduction}\label{sec:intro}

The collapsar model~\cite{Woosley1993, 1998ApJ...494L..45P, Macfadyen1999} is one of the most promising models for explaining the central engine of long gamma-ray bursts (GRBs) (see also Refs.~\cite{Piran2004, Zhang:2003uk} for a review). This model supposes the presence of a massive, rotating, and magnetized progenitor star that collapses into a spinning black hole surrounded by a massive torus. The black hole is often supposed to be penetrated by a poloidal magnetic field with a sufficiently high field strength $\agt 10^{14}$\,G, with which the Poynting luminosity by the Blandford-Znajek mechanism~\cite{Blandford1977} is high enough to explain the luminosity of long GRBs. The formed massive torus is supposed to be in a turbulent state due to the magnetohydrodynamical instability and the resulting heating by the effectively viscous process could be an engine for the stellar explosion~\cite{Kohri2005aug, Just2022aug, Fujibayashi:2023oyt, Dean_2024}, which is often associated with long GRBs~\cite{Cano2017a}. This model has stimulated a number of general relativistic magnetohydrodynamics simulations (in the fixed black-hole spacetime) in the last two decades (e.g., Refs.~\cite{Komissarov:2005wj,McKinney:2006tf,2008MNRAS.385L..28B,Komissarov:2009dn,Tchekhovskoy:2011zx,Bromberg:2015wra,Gottlieb:2021srg}), which indicated that jets are indeed launched in the presence of strong poloidal magnetic fields that penetrate a spinning black hole, which are hypothetically assumed. The jet could also be the source of the stellar explosion~(e.g., Refs.~\cite{2007ApJ...657L..77T, 2012ApJ...750...68L}), although the observational results  may not support this idea~\cite{Eisenberg2022nov}. 

In our previous paper~\cite{Shibata:2023tho}, we performed axisymmetric ideal-magnetohdydrodynamics simulations for modeling collapsar incorporating a neutrino-radiation transfer in full general relativity for the first time. In that work, we employed a stellar evolution model developed for collapsars~\cite{Aguilera-Dena2020oct} and focused on the stage after the formation of a spinning black hole. We assumed the presence of a poloidal magnetic field and investigated the evolution of the black hole and magnetic fields by the infalling stellar envelope taking fully into account the self-gravity of the matter infalling toward the central region. Simulations were performed for a timescale of 10--20\,s to follow the winding of the magnetic fields associated with the black-hole spin. We found the following results: (i) a jet \footnote{In this paper we refer to a collimated outflow propagated along the spin axis of black holes as a jet even if it is not ultra-relativistic.}
starts launching at the moment that the enhanced magnetic pressure exceeds the ram pressure of the infalling matter, i.e., when the condition 
\beq
{B^2 \over 8\pi} > \rho_\mathrm{inf} v_\mathrm{inf}^2 \label{eq1}
\eeq
is approximately satisfied in the vicinity of the spin axis of the black hole. Here $B$ denotes the magnetic-field strength in the polar region near the black hole, which is enhanced by the winding associated with the black-hole spin, and $\rho_\mathrm{inf}$ and $v_\mathrm{inf}$ are the rest-mass density and infall velocity of the matter; (ii) the magnetic-field strength on the horizon is approximately preserved after the onset of the jet launch (a magnetically arrested state~\cite{2003PASJ...55L..69N,Igumenshchev:2003rt,Tchekhovskoy:2011zx} is established), and the Poynting flux associated with the Blandford-Znajek mechanism remains approximately constant; (iii) as a result of the continuous extraction of the rotational kinetic energy of the black hole, its spin decreases with time, if the matter infall onto the black hole is negligible. For the models in which a jet is launched in an earlier stage, the magnetic-field strength is stronger because of a higher value of $\rho_\mathrm{inf}$, and the spin-down timescale can be shorter than or as short as the typical time duration of long GRBs $\sim 10$--$100$\,s. (iv) for such models, the total energy extracted from the black hole can be much larger than $10^{53}$\,erg, which is larger than the typical energy of long GRBs, its afterglow, and supernovae (SNe) often associated with them (see also Refs.~\cite{Corsi2016, 2024arXiv241112850W}). From these results, we concluded that it is not very likely that progenitor stars of collapsars have strong {\em poloidal} magnetic fields before the formation of an accretion disk around a black hole, and speculated that the strong poloidal magnetic field for the collapsars should be developed in the disk/torus surrounding the black hole by magnetohydrodynamics instabilities such as magnetorotational instability (MRI) and Kelvin-Helmholtz instability, which induce turbulent motion~\cite{Balbus:1998ja}, and by the subsequent accretion of poloidal magnetic fields accompanied with the matter infall onto the black hole. 

To verify this hypothesis, we have to follow the turbulent motion resulting from the magnetohydrodynamical dynamo process in the disk/torus resolving MRI (and other magnetohydrodynamical instabilities) with an initial condition of weak or negligible poloidal magnetic field. For this purpose, a three-dimensional simulation with the timescale of much longer than $10$\,s is required, if we rely on the ideal magnetohydrodynamics. Although such a simulation should be one of the goals in relativistic astrophysics, unfortunately, it is not feasible in the current computational resources. Thus, in this paper, following our previous paper~\cite{Shibata2021b}, we perform axisymmetric resistive magnetohydrodynamics simulations adding a dynamo term, which enables us to take into account the turbulence effect in the disk/torus phenomenologically. This method is not suitable for deriving quantitatively accurate results for the explosion energy and Poynting flux associated with the Blandford-Znajek mechanism. However, it is still useful for developing the qualitative and semi-quantitative picture for the mechanism of the jet launch and stellar explosion in the collapsar scenario. Furthermore, the assumption of the axisymmetry enables us to perform fully general relativistic simulations for a sufficiently long term, i.e., $30$--$50$\,s, which is necessary to clarify the mechanisms of the stellar explosion and jet launch self-consistently. 

It is also worth referring to the recent works of long-term simulations for neutron-star mergers in general relativity~\cite{Hayashi:2021oxy, Hayashi:2022cdq, Kiuchi:2022nin, Hayashi:2024jwt}. These works follow the long-term evolution of remnants of neutron star mergers, which are composed of a spinning black hole and accretion disk by general relativistic ideal magnetohydrodynamics simulations. They found that in the accretion disks, a turbulence is developed by the MRI and associated $\alpha$-$\Omega$ dynamo, which subsequently enhance the magnetic-field strength in the disk and induce mass ejection. Subsequent magnetic-flux accretion, associated with the matter accretion from the disk, onto the black hole constructs a magnetosphere near the spin axis of the black hole. For the case that the matter density near the spin axis is low, they also found the Poynting-flux-driven collimated outflow along the spin axis~\cite{Hayashi:2021oxy, Hayashi:2022cdq, Hayashi:2024jwt}. Hence, assuming the development of the turbulent state by the dynamo action in the disk/torus around a black hole, which can be the source of the subsequent powerful phenomena, is quite reasonable. 

In the present setup, we indeed find that a poloidal magnetic field that penetrates the spinning black hole is formed and a jet may be launched after the development of a torus in a turbulent state. This turbulent state of the torus also becomes the engine for the entire stellar explosion with the explosion energy as high as or higher than typical SNe of $10^{51}$\,erg~\cite{Fujibayashi:2023oyt}. Thus, in this scenario, long GRBs that accompany powerful SNe are self-consistently explained (see Sec.~\ref{sec4E} for details).

The paper is organized as follows. In Sec.~\ref{sec2}, we describe the setup of the present numerical simulations. In Sec.~\ref{sec3}, numerical results are shown focusing on the mechanisms of the stellar explosion and jet launch as well as on the energy of the stellar explosion and Poynting flux. We also perform a nucleosynthesis calculation to show that the explosion accompanies a large amount of $^{56}$Ni production as found in our viscous hydrodynamics simulations~\cite{Fujibayashi:2023oyt} and that jet launching can accompany a weak $r$-process nucleosynthesis~\cite{Issa:2024sts}. Section~\ref{sec4} is devoted to a summary and discussion. Throughout this paper, we use the geometrical units of $c=1=G$ where $c$ and $G$ denote the speed of light and gravitational constant, respectively. $k_\mathrm{B}$ denotes Boltzmann's constant.

\section{Set up}\label{sec2}

\begin{table*}[t]
\centering
\caption{Parameters for each model. Model name, initial maximum magnetic-field strength, the values of $\alpha_\mathrm{d}$, $\sigma_\mathrm{c}$, $\rho_\mathrm{cut}$, and the grid spacing in the central region $\Delta x$. 
The last two columns show whether we find the stellar explosion and jet launch. The last four rows list the data for high-resolution runs and a viscous hydrodynamics run. The model name refers to log$_{10}B_\mathrm{max}$, $10^4\alpha_\mathrm{d}$, log$_{10}\sigma_\mathrm{c}$, and high or low value of $\rho_\mathrm{cut}$. 
    \label{tab1}}
    \begin{tabular}{lccccccc}
    \hline
        Model & ~~$B_\mathrm{max}$ (G)~~ & ~~$\alpha_\mathrm{d}$~~ & ~$\sigma_\mathrm{c}$\,(s$^{-1}$)~ & ~$\rho_\mathrm{cut}$\,(g/cm$^{3}$)~ & ~~$\Delta x$\,(m)~~& 
        Explosion & Jet \\
        \hline
    B11.1.8h  & $10^{11}$ & $10^{-4}$  & $10^8$ & $10^8$ & 360 & Yes & Yes \\  
    B12.1.8h & $10^{12}$ & $10^{-4}$  & $10^8$ & $10^8$ & 360 & Yes & Yes \\   
    B12.1.8l & $10^{12}$ & $10^{-4}$  & $10^8$ & $10^6$ &360  & Yes & Yes \\   
    B12.3.8l & $10^{12}$ & $3\times 10^{-4}$  & $10^8$ & $10^6$ &360 & Yes & Yes \\ 
    B12.1.7l & $10^{12}$ & $10^{-4}$  & $10^7$ & $10^6$ &360 & Yes & Weak \\   
    B12.1.9l & $10^{12}$ & $10^{-4}$  & $10^9$ & $10^6$ &360 & Yes & Yes \\   
    \hline
B12.1.8l-H   & $10^{12}$   & $10^{-4}$ &  $10^8$ & $10^6$ &300 & Yes & Yes \\ 
B12.3.8l-H   & $10^{12}$   & $3\times 10^{-4}$ & $10^8$  & $10^6$ &300 & Yes & No \\ 
B12.1.7l-H   & $10^{12}$   & $10^{-4}$ & $10^7$  & $10^6$  &300 & Yes & Weak  \\ 
\hline    
viscous & --- & --- & --- & ---& 360 & Yes & No \\\hline
\end{tabular}\\
\end{table*}

We employ the same formulation and simulation code as in Refs.~\cite{Shibata:2021bbj,Shibata2021b} for the present neutrino-radiation resistive magnetohydrodynamics study. Specifically, we numerically solve neutrino-radiation resistive magnetohydrodynamics equations with a dynamo term in full general relativity in this code. A tabulated equation of state, DD2~\cite{banik2014a}, is employed, with the extension of the table down to low-density ($\rho\approx\SI{0.17}{g/cm^3}$) and low-temperature ($k_\mathrm{B}T=10^{-3}$\,MeV) region; see Ref.~\cite{Hayashi:2021oxy} for the procedure. We note that in the present context, in which the matter density is always lower than the nuclear-saturation density, the high-density part of the equation of state does not play an important role. Neutrino radiation transfer is incorporated using the leakage plus M1 transport scheme~\cite{Sekiguchi2015a,fujibayashi2017a}. This code can take into account the neutrino pair annihilation process approximately, but in the present context, this is a subdominant effect compared with the magnetohydrodynamics effects, so we switch it off. The electromagnetic energy, explosion energy, ejecta mass, and Poynting flux 
are evaluated by the same methods as in our previous paper~\cite{Shibata:2023tho}. 

Following Ref.~\cite{Shibata2021b}, the simulations are performed incorporating the terms which phenomenologically excite the $\alpha$-$\Omega$ dynamo. The dynamo action is determined by a dimensionless dynamo term $\alpha_\mathrm{d}$ and a conductivity $\sigma_\mathrm{c}$. In this paper, we employ $\alpha_\mathrm{d}=10^{-4}$ or $3 \times 10^{-4}$ and $\sigma_\mathrm{c}=10^k\,\mathrm{s}^{-1}$ with $k=7$, 8, and 9, assuming that the turbulence and dynamo are excited by the MRI (see the discussion in Ref.~\cite{Shibata2021b}; also, e.g.,  Refs.~\cite{2015ApJ...810...59G, 2021A&A...645A.109R, Kiuchi:2023obe} for a plausible value of $\alpha_\mathrm{d}$). We note that for higher values of $\alpha_\mathrm{d}$ the amplification of the magnetic field proceeds more quickly and for lower values of $\sigma_\mathrm{c}$ the dissipation of the magnetic field proceeds more quickly. 
In addition to these parameters, we introduce a cutoff density, $\rho_\mathrm{cut}$, to suppress the dynamo action in a low-density region, in which the magnetic pressure is comparable to or larger than the gas pressure (i.e., for magnetosphere). Specifically, the dynamo coefficient is modified as
\beq
\alpha_\mathrm{d} \rightarrow 
\alpha_\mathrm{d} [1-\exp(-\rho/\rho_\mathrm{cut})],
\eeq
where $\rho_\mathrm{cut}$ is primarily chosen to be $10^6$\,g/cm$^3$ and partly $10^8$\,g/cm$^3$ (see Table~\ref{tab1} for the parameters of each model). 




Following our recent work~\cite{Fujibayashi:2023oyt}, for the initial condition, we prepare a system of a spinning black hole and infalling matter instead of using the original progenitor star model. 
To obtain the initial data, specifically,  we take the progenitor models from a stellar evolution calculation of Ref.~\cite{Aguilera-Dena2020oct}. In this work we employ the $35M_\odot$ star model in their paper (\texttt{AD35} model of Ref.~\cite{Fujibayashi:2023oyt}), which is very compact at the onset of the stellar core collapse, and hence, it is reasonable to assume that a black hole is formed in a short timescale after the onset of the core collapse~\cite{Oconnor2011apr}. Since the angular momentum in the inner region of the progenitor star is not very large, the black hole should be evolved by the accretion from the outer region without forming an accretion disk for a while~\cite{Fujibayashi:2023oyt}. We select such a stage to construct the initial data by solving constraint equations of general relativity in the hypothesis that the system is momentarily composed of a spinning black hole and nearly free-falling matter. Since we set up the initial data at a stage prior to the formation of a disk, the mass and dimensionless spin of the black hole are high as $M_\mathrm{BH,0}=16M_\odot$ and $\chi_0=0.70$, and the rest mass outside the black hole is $M_\mathrm{mat}\approx 9.5M_\odot$. Here, the initial angular momentum of the black hole is $J_\mathrm{BH,0}=M_\mathrm{BH,0}^2\chi_0$. 

The mass and dimensionless spin of the black hole are determined by analyzing the equatorial and polar circumferential radii, $C_e$ and $C_p$, respectively, of apparent horizons~(e.g., see Ref.~\cite{Shibata2016a}). Specifically, the mass is determined by the relation of 
\beq
M_\mathrm{BH}= {C_e \over 4\pi},
\eeq
and the dimensionless spin is determined from $C_p/C_e$, which is a monotonic function of the dimensionless spin, $\chi$, for Kerr black holes and can be used to identify the value of $\chi$. We also confirm that the mass and spin obtained by them satisfy the relation of the area, $A_\mathrm{AH}=8\pi M_\mathrm{BH}^2(1 + \sqrt{1-\chi^2})$, within the error of 0.1\%. 
We cut out the matter outside the radius of $10^5$\,km because the computational domain in our simulation is $10^5\times10^5$\,km for $\varpi$ and $z$ where $\varpi$ is the cylindrical coordinate. 

We superimpose a toroidal magnetic field with which the electromagnetic energy is initially much smaller than the internal energy and kinetic energy. Specifically we give
\beqn
B^y=B_0\sqrt{\mathrm{max}(P-10^{-20}P_\mathrm{max},0)}\,f(x)f(|z|), \label{initB}
\eeqn
where $P$ denotes the gas pressure with $P_\mathrm{max}$ its maximum, $y$ is the toroidal direction in the $x$-$z$ plane, $B_0$ is a constant, and
\beq
f(x)={2 \over e^{-x/R_0}+1}-1, 
\eeq
where we set $R_0=10$\,km. Note that $f(x)f(|z|)$ is necessary to guarantee the regularity relation of the magnetic field at the symmetric axis and on the equatorial plane (we impose the equatorial-plane symmetry in the simulation). The initial magnetic field strength is controlled by the value of $B_0$, and in this work, we choose it so that the maximum field strength becomes $10^{12}$ or $10^{11}$\,G. With the initial condition of Eq.~(\ref{initB}) the ratio of the magnetic pressure to the gas pressure is approximately constant except for the region near the symmetric axis and equatorial plane. 

Since the magnetic pressure is much weaker than the gas pressure in the present setting, the magnetic field effects play no role before the dense disk/torus is established. Also, because of the absence of the initial poloidal magnetic field, the Blandford-Znajek mechanism is negligible in the early stage of the evolution of the system, i.e., before the development of a turbulence in the disk/torus. The magnetic field effect including the Blandford-Znajek mechanism plays an important role only after a turbulent state of the disk/torus is established. 

As discussed in our previous paper~\cite{Shibata:2023tho} the magnetic-field strength required on the horizon is $B\sim 10^{14}$\,G for typical long GRB jets because the luminosity in the Blandford-Znajek mechanism is written as
\beqn
    L_\mathrm{BZ}
    &\approx & 1 \times 10^{50}\,
    \left({M_{\rm BH} \over 10M_\odot}\right)^2 
    \left({B_\mathrm{p} \over 10^{14}\,{\rm G}}\right)^2
    \left({\chi \over 0.7}\right)^2\,{\rm erg/s}, \nonumber \\\label{eq10}
\eeqn
where $B_\mathrm{p}$ is the poloidal magnetic-field strength on the horizon while $B$ denotes the total field-strength. 
The magnetic pressure for such a field strength is $B^2/8\pi=O(10^{26})$\,${\rm dyn/cm}^2$, which has to be larger than the ram pressure of the infalling matter at the jet launch. For given values of the rest-mass density $\rho_\mathrm{inf}$ and the infall velocity $v_\mathrm{inf}$, the ram pressure is written as
\beqn
\rho v_\mathrm{inf}^2
&\approx &2.2 \times 10^{26}\left({\rho_\mathrm{inf} \over 10^6\,{\rm g/cm^3}}\right)
\left({v_\mathrm{inf} \over c/2}\right)^2\,{\rm dyn/cm^2}.~~~
\label{eq:ram1}
\eeqn
Thus a jet is likely to be launched in a late stage of the system in which the density of the infalling matter decreases below $\sim 10^6\,{\rm g/cm^3}$. 

The computational region is prepared in the same manner as in Ref.~\cite{Shibata:2023tho}. The simulation is performed on a two-dimensional domain of $\varpi$ and $z$ (see also Refs.~\cite{Fujibayashi2020a,Fujibayashi2020b}). For the $\varpi$ and $z$ directions, a non-uniform grid is employed: For $x \alt 7GM_\mathrm{BH,0}/4c^2$ ($x=\varpi$ or $z$), a uniform grid is used, while outside this uniform region, the grid spacing $\Delta x_i$ is increased uniformly as $\Delta x_{i+1}=1.01\Delta x_i$, where the subscript $i$ denotes the $i$-th grid. 
The black-hole horizon (apparent horizon) is always located in the uniform grid zone, and the outer boundaries along the $\varpi$ and $z$ axes are located at $\approx 10^5$\,km. The grid resolution of the uniform grid zone is $\Delta x =360\,\mathrm{m} \approx 0.0152GM_\mathrm{BH,0}/c^2$ for the standard runs (see Appendix B of Ref.~\cite{Fujibayashi:2023oyt} for the validity of this choice). For three models (B12.1.7l, B12.1.8l, and B12.3.8l) higher resolution runs with $\Delta x =300\,\mathrm{m}\approx 0.0127GM_\mathrm{BH,0}/c^2$ is performed to examine the dependence of the numerical results on the grid resolution. A high-resolution simulation was also performed for $\alpha_\mathrm{d}=3 \times 10^{-4}$ and $\sigma_\mathrm{c}=10^7\,\mathrm{s}^{-1}$ (referred to as B12.3.7l-H in the following), but the result is similar to model B12.3.8l-H, and hence, we do not discuss the results in this paper. We also perform a viscous hydrodynamics simulation for comparison with the same setup as the magnetohydrodynamics simulation with grid resolution of 360\,m. In this paper we start the viscous hydrodynamics simulation with $M_\mathrm{BH,0}=16M_\odot$ (in Ref.~\cite{Fujibayashi:2023oyt} $M_\mathrm{BH,0}=15M_\odot$) and choose the dimensionless alpha parameter as $\alpha_\nu=0.03$ 
(see Ref.~\cite{Fujibayashi:2023oyt} for the definition of it). Table~\ref{tab1} lists the set-up information for all the models discussed in this paper. 

For each model, the system was evolved for 30--50\,s. Each run was performed on the Sakura or Momiji clusters at the Max Planck Computing and Data Facility. Typical computational costs were 3 million cpu hours. 

\section{Numerical results}\label{sec3}

\begin{figure*}[t]
\includegraphics[width=0.325\textwidth]{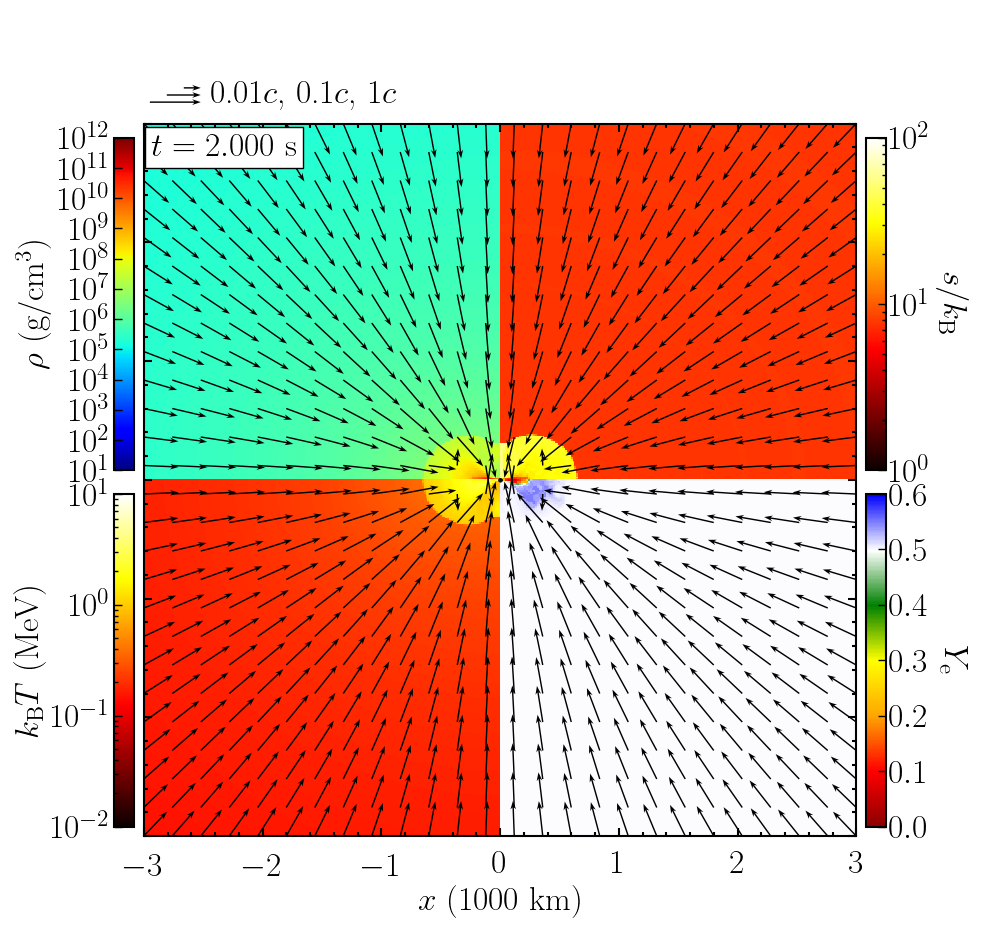}
\includegraphics[width=0.325\textwidth]{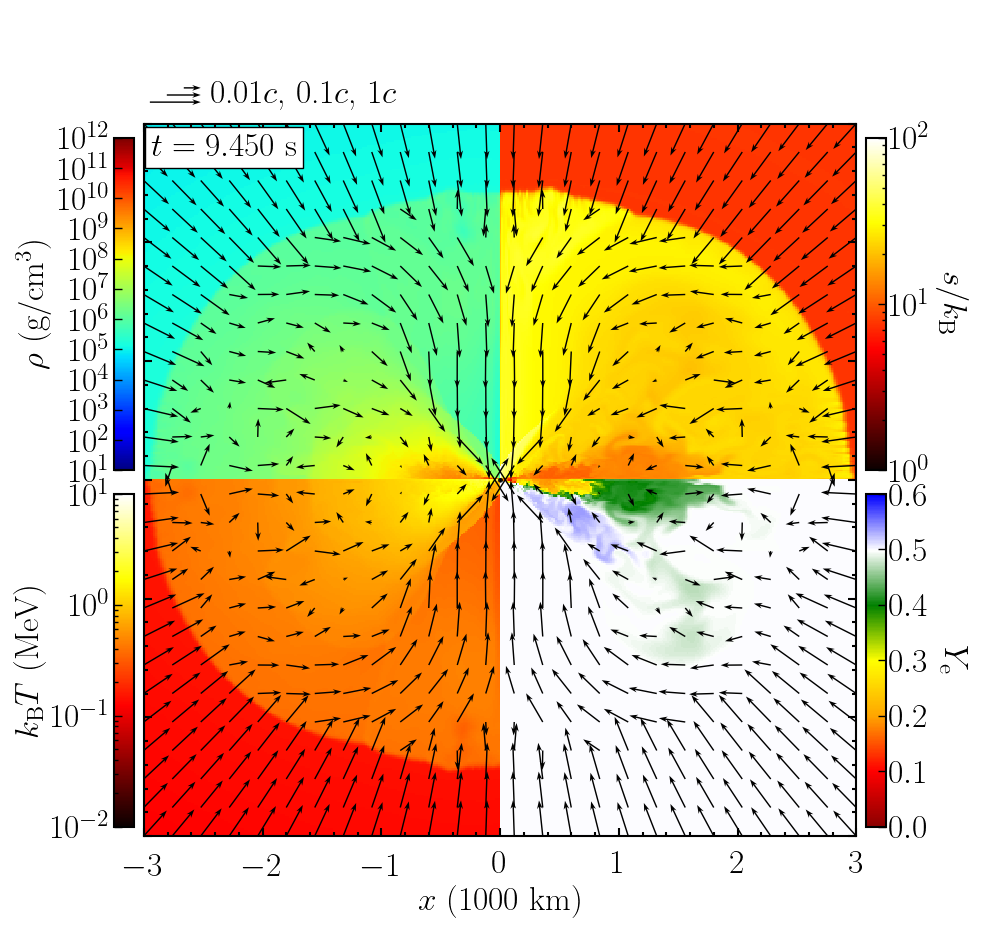}
\includegraphics[width=0.325\textwidth]{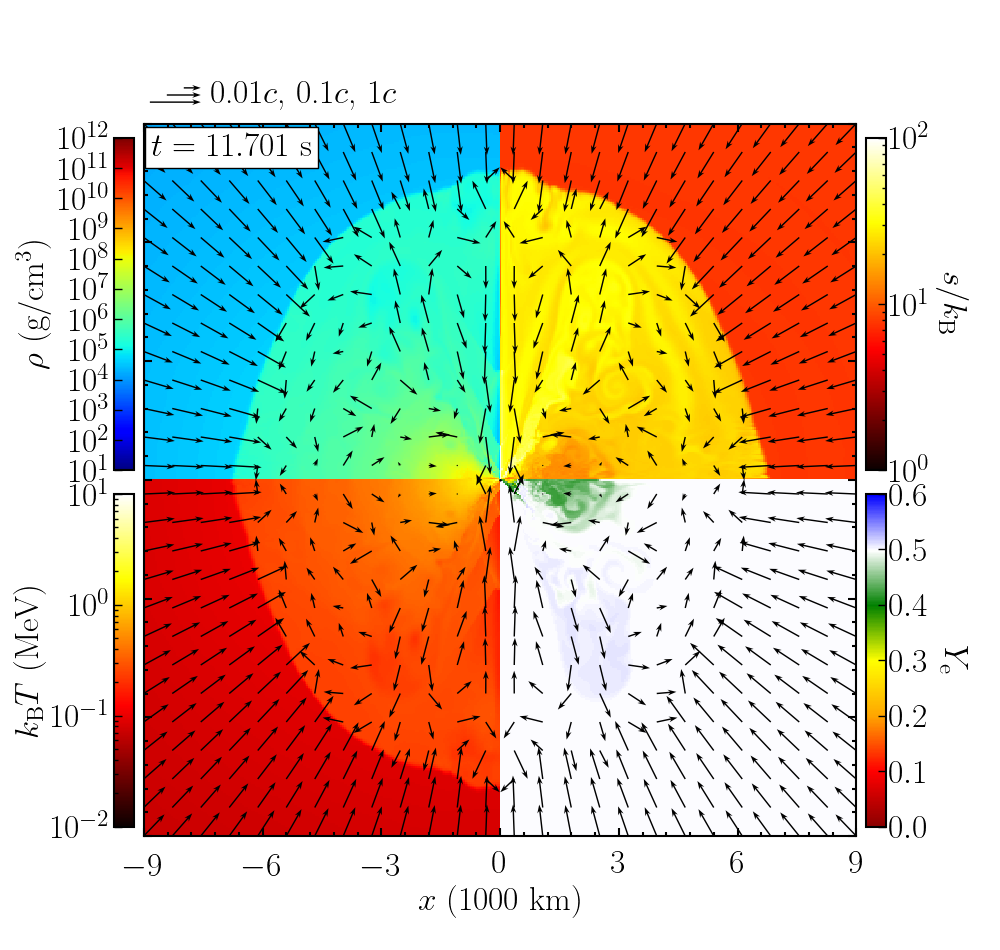}\\
\includegraphics[width=0.325\textwidth]{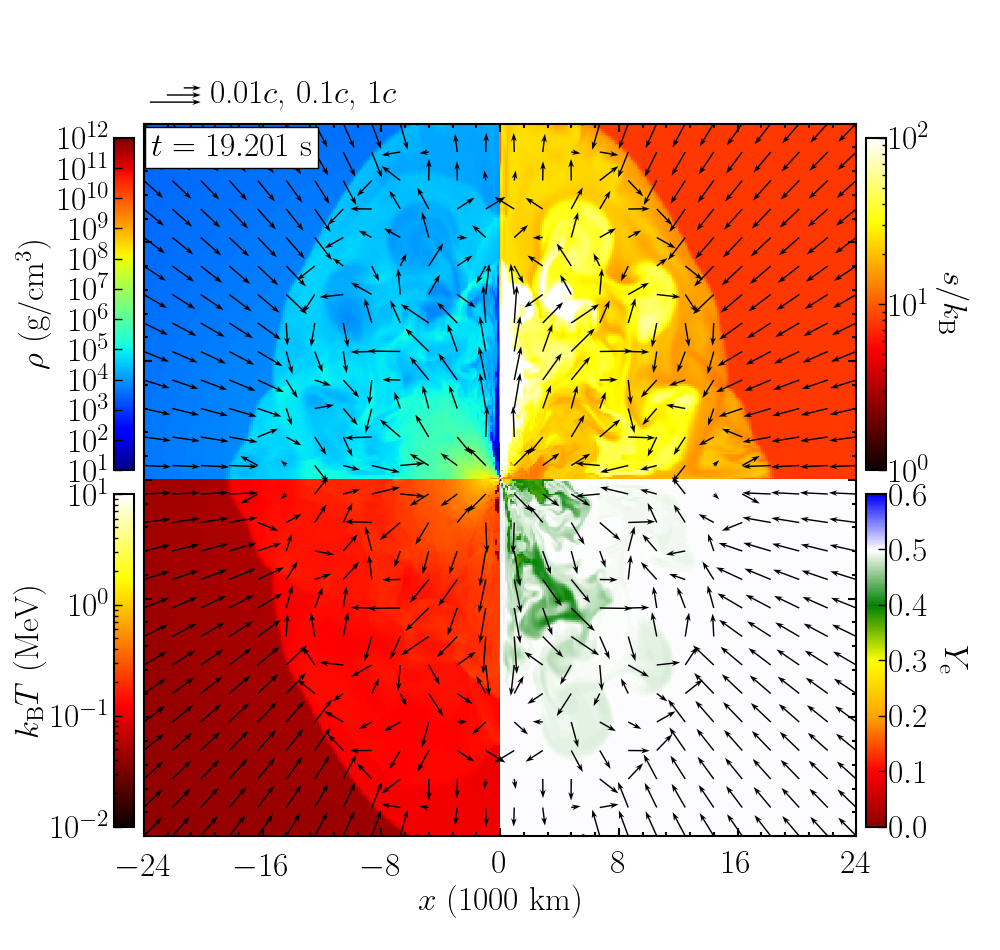}
\includegraphics[width=0.325\textwidth]{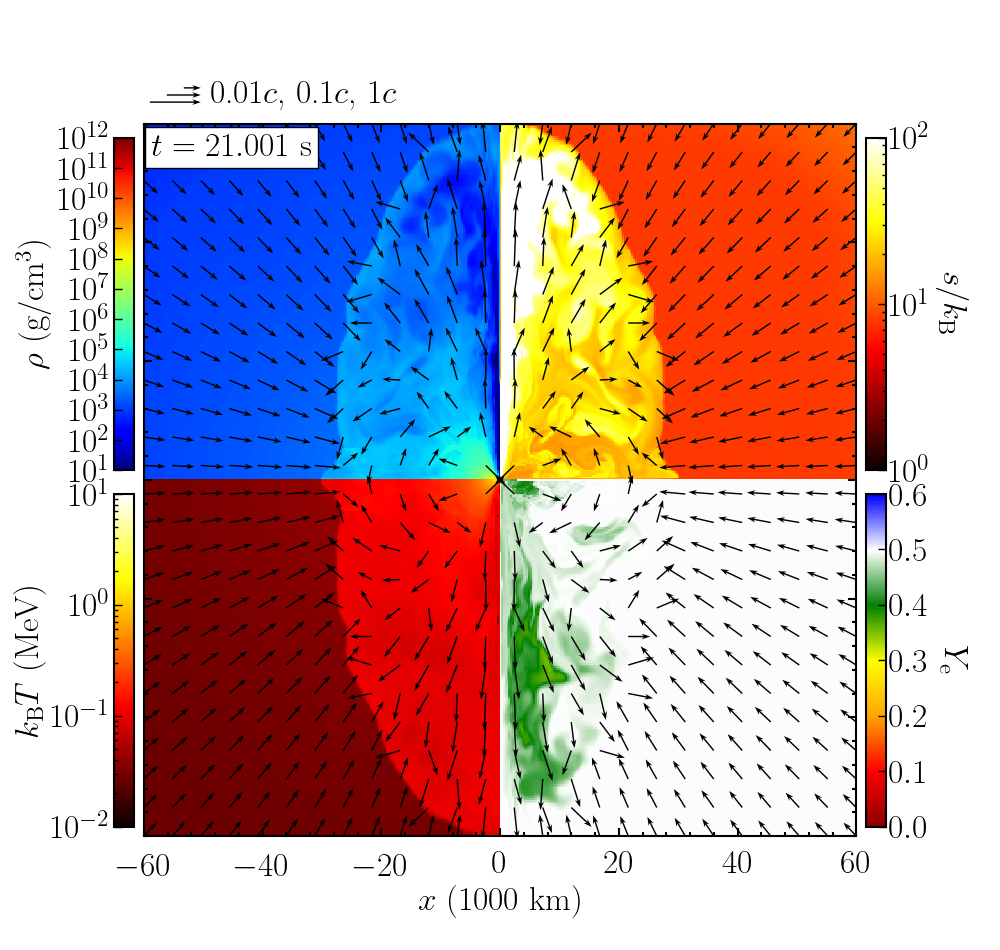}
\includegraphics[width=0.325\textwidth]{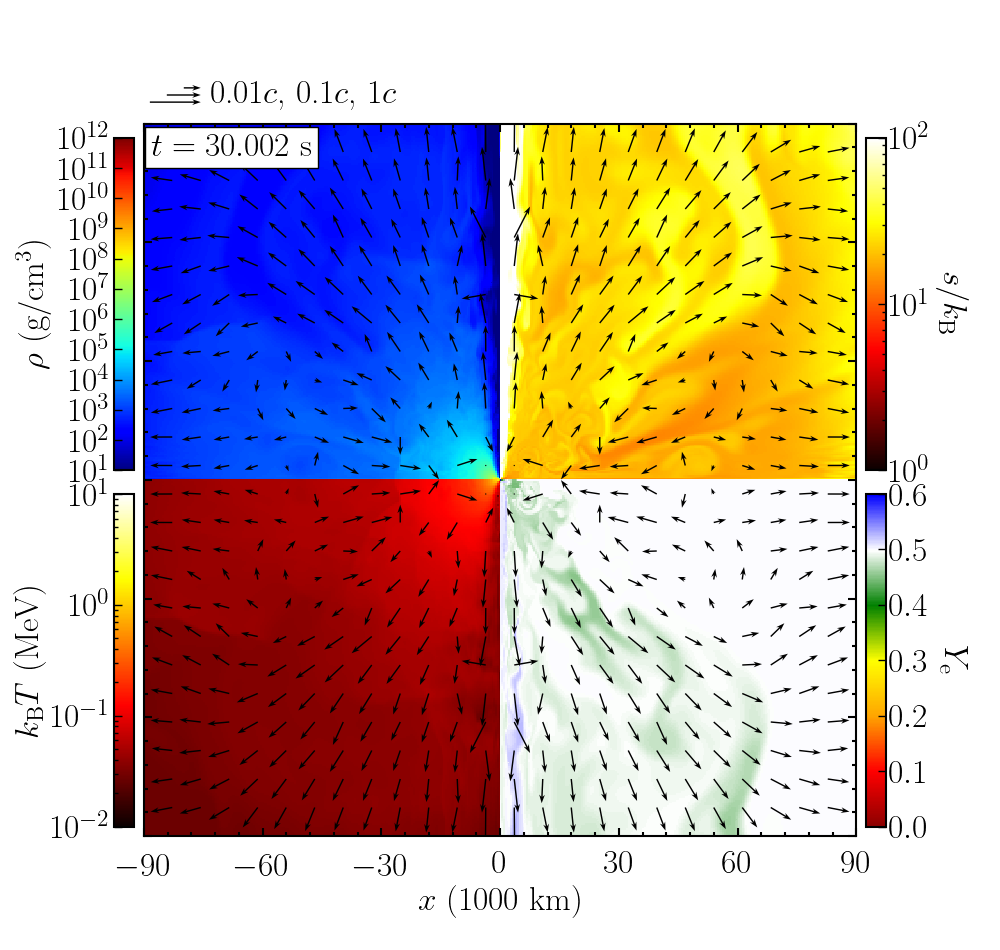}
\caption{Snapshots of the rest-mass density (top-left), entropy per baryon (top-right), temperature (bottom-left), and electron fraction (bottom-right) on the $\varpi$-$z$ plane at selected time slices for model B11.1.8h. Note that for each snapshot (except for the first two snapshots), the regions displayed are different. An animation for this model is found at \url{https://www2.yukawa.kyoto-u.ac.jp/~sho.fujibayashi/share/B11.1.8h-multiscale.mp4}
}
\label{fig1}
\end{figure*}

\begin{figure*}[t]
\includegraphics[width=0.35\textwidth]{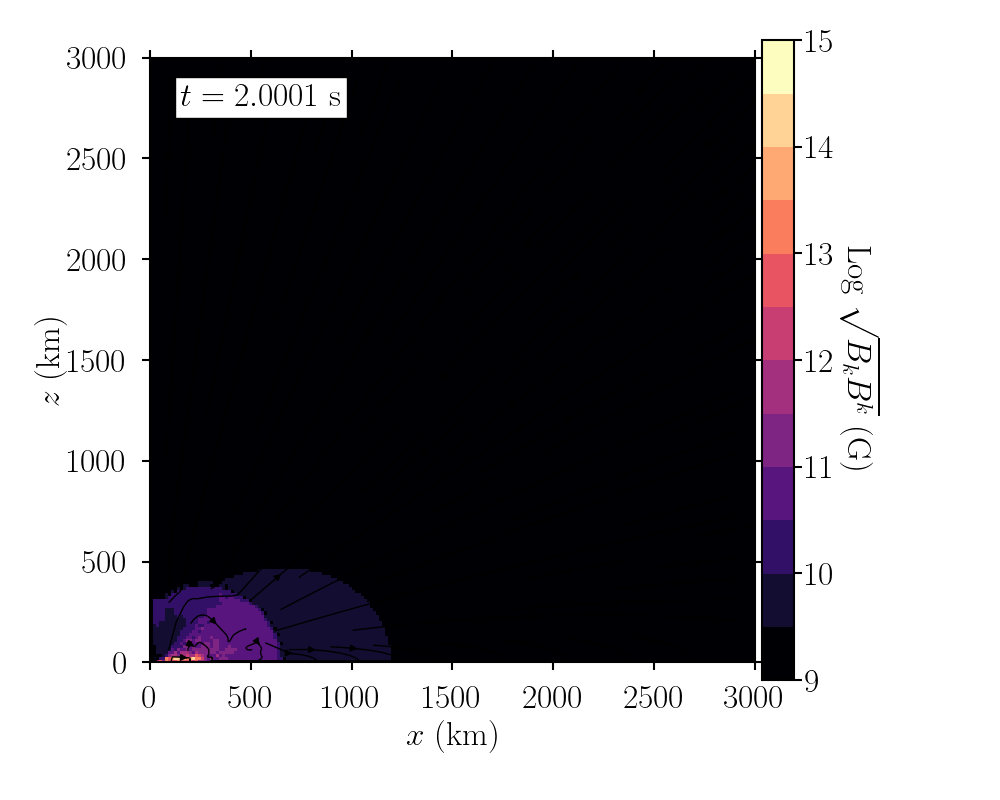}
\hspace{-7mm}
\includegraphics[width=0.35\textwidth]{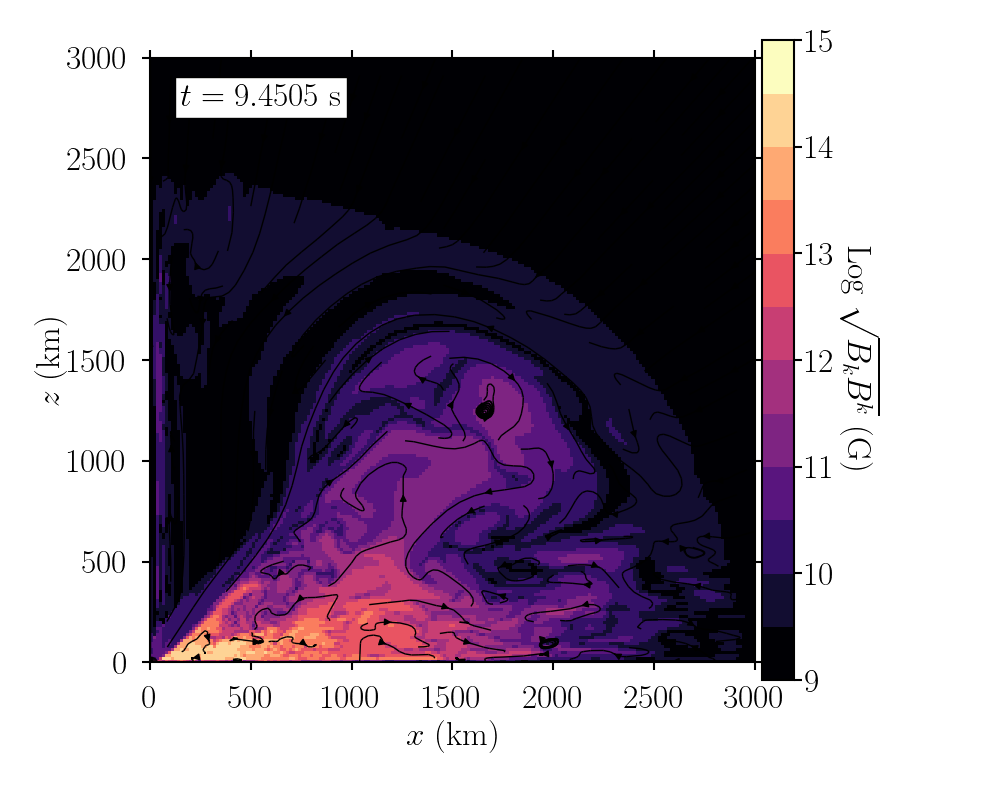}
\hspace{-7mm}
\includegraphics[width=0.35\textwidth]{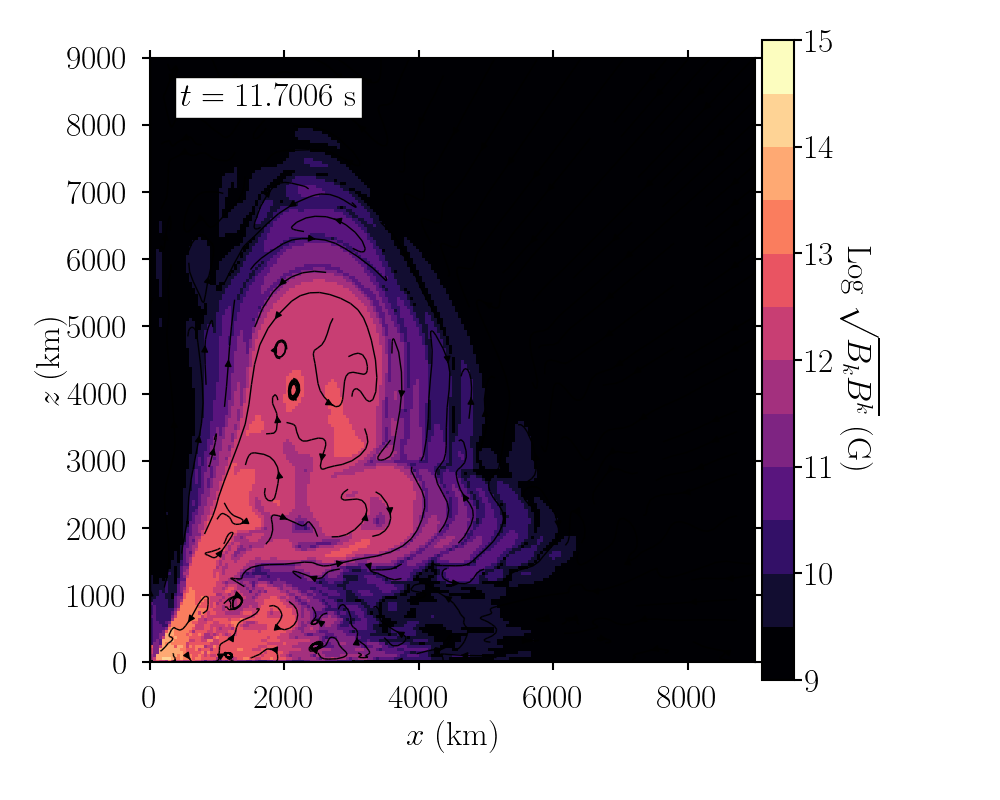}\\
\includegraphics[width=0.35\textwidth]{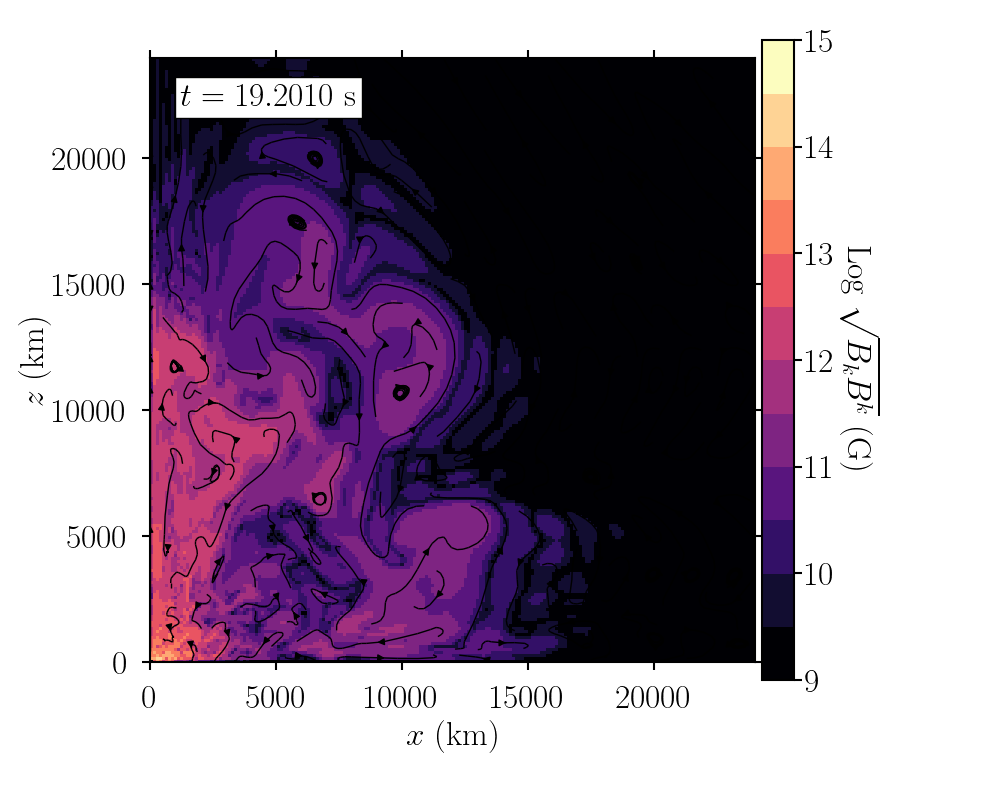}
\hspace{-7mm}
\includegraphics[width=0.35\textwidth]{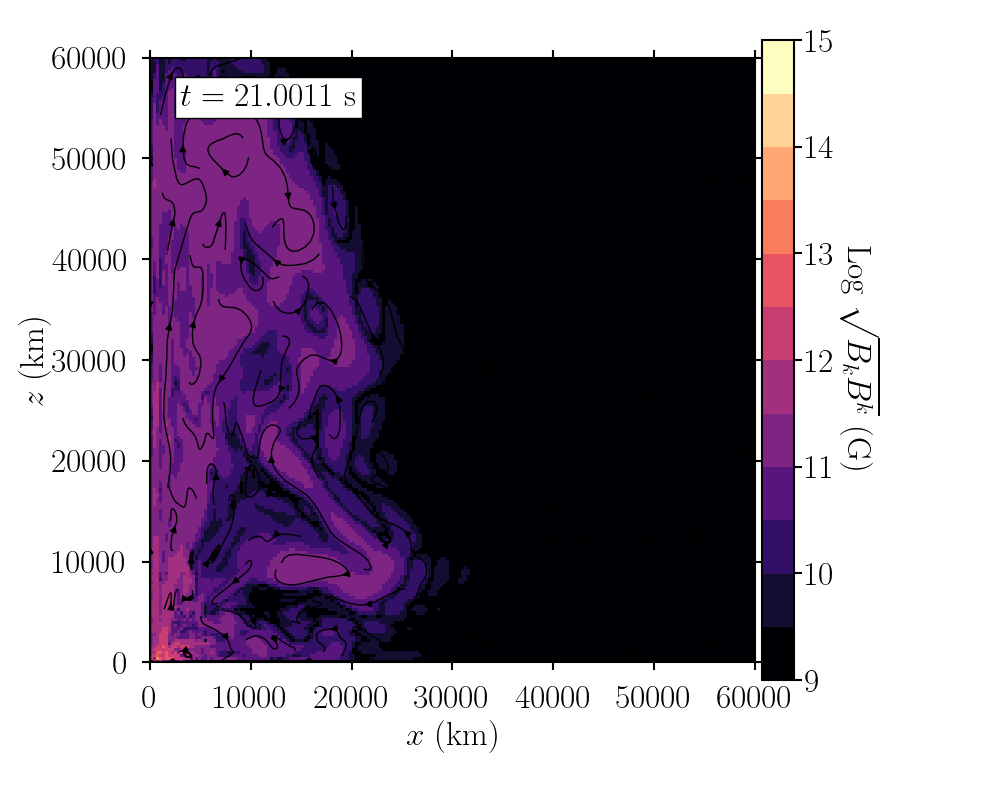}
\hspace{-7mm}
\includegraphics[width=0.35\textwidth]{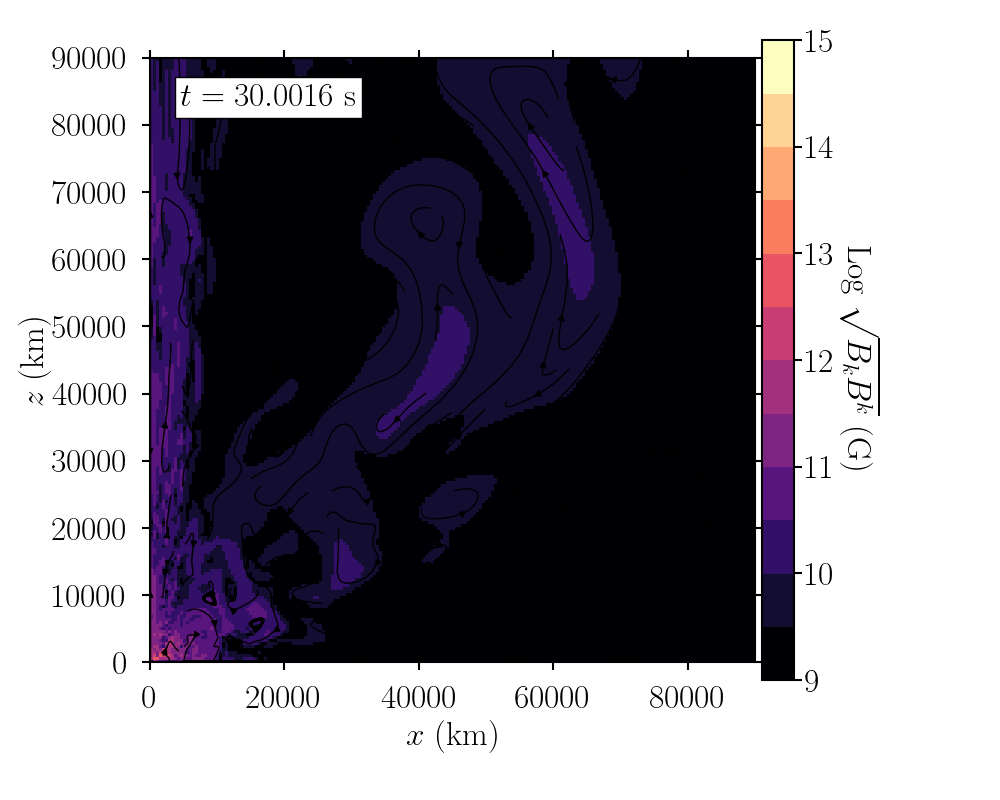}
\caption{The same as Fig.~\ref{fig1} but for the magnetic-field strength (color) and poloidal field lines. An animation for this model is found at \url{https://www2.yukawa.kyoto-u.ac.jp/~sho.fujibayashi/share/B11.1.8h-multiscale-B.mp4}
}
\label{fig2}
\end{figure*}

\subsection{Mechanism of stellar explosion and jet launch}


For all of the models simulated in this paper, we find the stellar explosion although the energy of the stellar explosion depends strongly on the parameters $(\alpha_\mathrm{d}, \sigma_\mathrm{c})$ chosen. The stellar explosion is driven primarily by the magnetohydrodynamics activity of the torus. Jet launch is also found for many of the models but for some of the models, we do not find the establishment of a strong jet along the axis of the black hole spin (see Table~\ref{tab1} and the discussion below for the reason for this). 
We here first summarize the evolution process until the stellar explosion and jet launch describing the entire evolution process for model B11.1.8h (see Figs.~\ref{fig1} and \ref{fig2}), for which we find both the successful stellar explosion and jet launch. We note that for other models in which the explosion and jet launch are found, the evolution process is qualitatively identical. 

\begin{figure*}[t]
\includegraphics[width=0.495\textwidth]{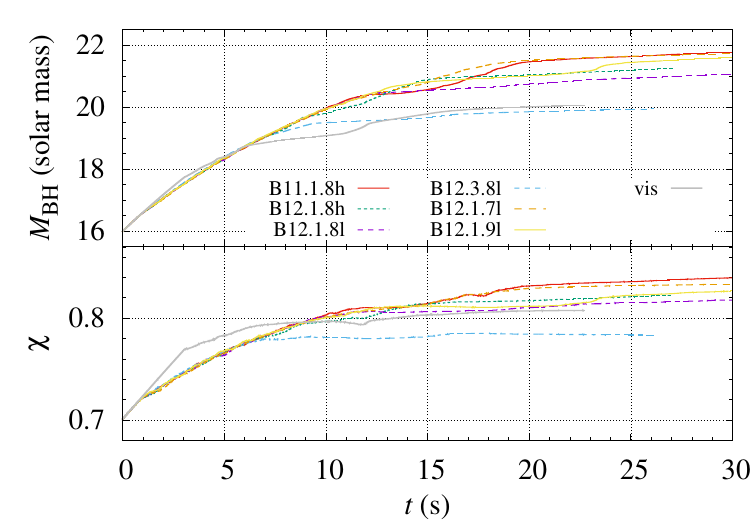}
\includegraphics[width=0.495\textwidth]{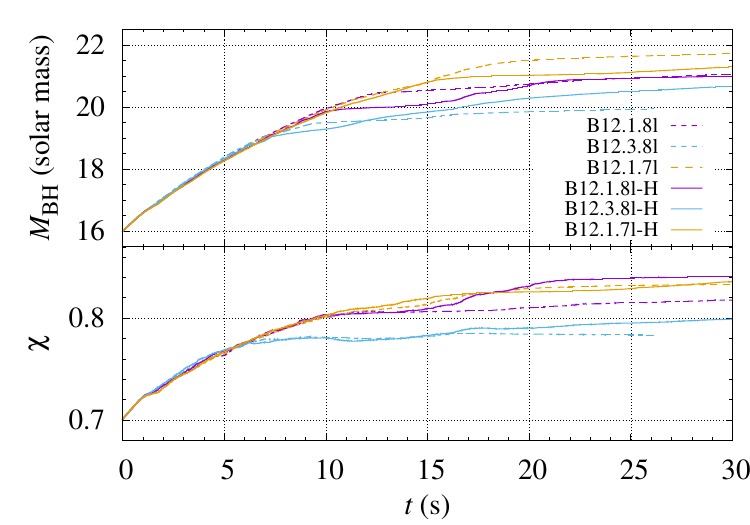}
\caption{Left: Evolution of the mass and dimensionless spin of the black holes for models B11.1.8h, B12.1.8h, B12.1.8l, B12.3.8l, B12.1.7l, and B12.1.9l as well as for a viscous hydrodynamics model. Right: The same as the left panel but for models B12.1.8l, B12.3.8l, B12.1.7l (dashed curves), and their high-resolution runs (solid curves). 
}
\label{fig3}
\end{figure*}

\begin{figure*}[t]
\includegraphics[width=0.495\textwidth]{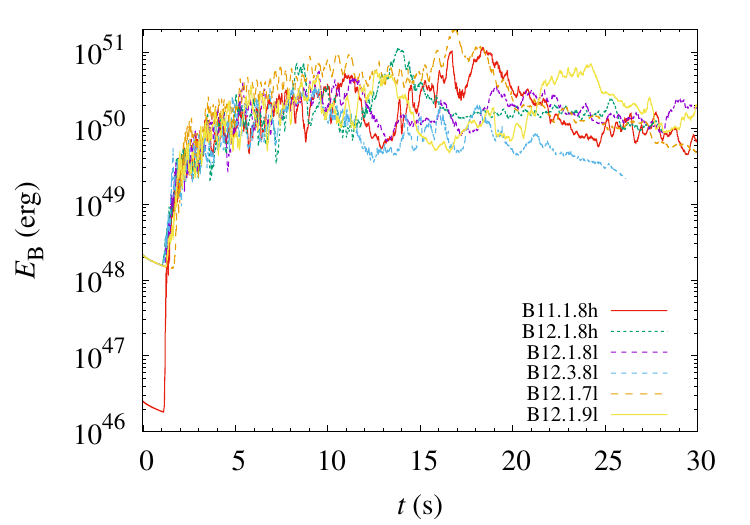}
\includegraphics[width=0.495\textwidth]{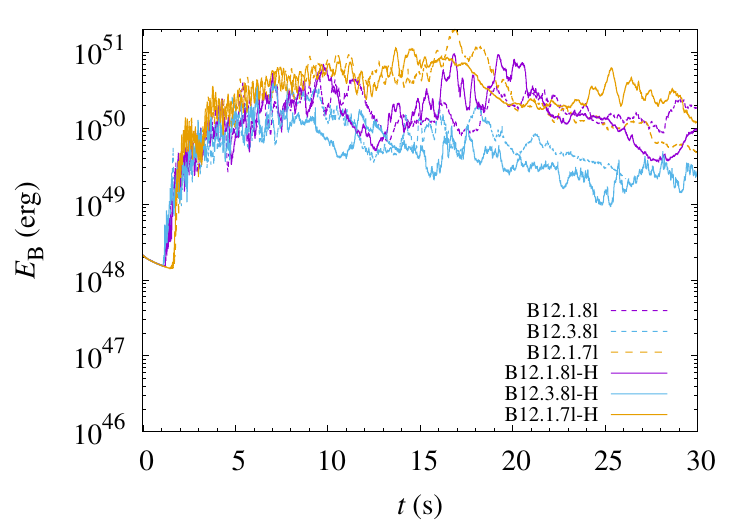}
\caption{Left: Evolution of the electromagnetic energy for models B11.1.8h, B12.1.8h, B12.1.8l, B12.3.8l, B12.1.7l, and B12.1.9l. Right: The same as the left but for models B12.1.8l, B12.3.8l, B12.1.7l (dashed curves), and their higher-resolution runs (solid curves).
}
\label{fig4}
\end{figure*}

For the first $\sim 1.5$\,s after the start of the simulation, the mass infall onto the black hole simply proceeds with no disk formation, and as a result, the mass and dimensionless spin of the black hole monotonically increase with time (cf. Fig.~\ref{fig3}). The situation changes at $t \sim 1.5$\,s, at which the disk and subsequently geometrically thick torus are developed around the black hole due to the infall of the matter with relatively high specific angular momentum (see the second and third panels of Fig.~\ref{fig1}), although the mass infall onto the black hole still proceeds steadily from the polar region. The total mass and radial extent of the torus as well as the mass and spin of the black hole monotonically increase with time spending $\sim 10$\,s, and then, due to the dynamo action, a turbulent state is developed in the torus, enhancing the magnetic-field strength (see Fig.~\ref{fig4} for the growth and saturation of the electromagnetic energy). However, in the early stage of the torus evolution, i.e., for the first $\sim 10$\,s for model B11.1.8h, the ram pressure of the infalling matter is still larger than the electromagnetic forces, and hence, the expansion of the torus is suppressed although the torus gradually expands in the radial direction due to the angular momentum transport as well as the infall of the matter with high specific angular momentum from the outer envelope. We note that all these processes universally proceed for all the models studied in this paper. 

After the turbulent state is developed in the torus, the mass accretion from the torus onto the black hole also proceeds due to the associated angular momentum transport process. By this process, magnetic fluxes fall onto the black hole, and hence, the poloidal magnetic fields that penetrate the black hole could be developed. Our numerical simulations show that the typical field strength in the torus is determined by the equi-partition relation between the magnetic energy density and internal energy density, which is approximately written as $B^2/8\pi=\eta\rho c_\mathrm{s}^2$ where $\rho$ denotes the typical density of the torus, $c_\mathrm{s}$ the typical sound velocity, and $\eta$ is a constant of 0.01--0.1. In the vicinity of the black hole, the typical density and sound velocity are $10^9$--$10^{10}\,\mathrm{g/cm^3}$ and $0.1c (=3 \times 10^9\,\mathrm{cm/s})$ before the stellar explosion and jet launch. Then we have
\beqn
B&=&1.1 \times 10^{14}\,\mathrm{G} \left({\rho \over 10^{9}\,\mathrm{g/cm^3}}\right)^{1/2}
\left({c_\mathrm{s} \over 3 \times 10^9\,\mathrm{g/cm^3}}\right)
\nonumber \\
&&~~~~~~~~~~~~~~\times\left({\eta \over 0.05}\right)^{1/2}. \label{eq8}
\eeqn
This field strength is a good number to produce typical GRBs if the poloidal magnetic field that penetrates the black hole is developed through the magnetic-flux accretion from the torus (see Eq.~(\ref{eq10})) and if the conversion efficiency of the Poynting flux to gamma-rays is sufficiently high. 

In the early stage of the evolution (for model B11.1.8h, until $\sim 19$\,s), the ram pressure by the infalling matter is still higher than the electromagnetic force near the spin axis of the black hole, and thus, a seed of the poloidal magnetic field is swallowed by the black hole before a sufficient amplification by the winding associated with the black-hole spin. In such a situation, it is not possible to construct a robust large-scale magnetosphere that can be responsible for the Blandford-Znajek mechanism. We note that the duration for this stage i.e., until the establishment of a large-scale magnetosphere, depends strongly on the dynamo parameters and cutoff density. 

However, the ram pressure decreases with time due to the decrease of the density of the infalling matter. As a result, the electromagnetic forces can eventually overcome the ram pressure (cf.~Eq.~(\ref{eq1})), leading to a mass ejection from the torus and to a jet launch (see the third and fourth panels of Figs.~\ref{fig1} and \ref{fig2}). The mass ejection power of the torus is prominent in particular from its inner edge (close to the black hole) for which the electromagnetic force is strongest. This mass ejection power is not still large enough to induce the stellar explosion in an early stage, but the expansion of the torus is accelerated by this activity. The expansion of the torus can block the mass accretion onto the black hole from the envelope, and thus, by this activity, the ram pressure near the black hole is reduced. Under such a situation, if a poloidal magnetic field, which is supplied from the torus, is amplified by the winding associated with the black-hole spin and the resulting electromagnetic force overcomes the ram pressure of the infalling matter, a jet-like outflow is driven from the polar surface of the black hole (see the fourth panel of Fig.~\ref{fig1}). After the launch of the outflow, a funnel structure with low density and high entropy per baryon is established in the polar region (see the fourth--sixth panels of Fig.~\ref{fig1}), and the Poynging flux associated with the Blandford-Znajek mechanism starts being propagated outwards (see the fourth panel of Fig.~\ref{fig2}), leading to the development of the jet structure near the spin axis. 

Approximately at the same time, entire stellar explosion also sets in for model B11.1.8h (and for other jet-launching models). This results from the magnetohydrodynamics activity and strong shock heating at the inner region of the torus close to the black hole (see fifth and sixth panels of Figs.~\ref{fig1} and \ref{fig2}). The magnetocentrifugal force associated with the black hole spin and differential rotation of the torus may partly contribute to the mass ejection from the torus. The jet developed near the spin axis also contributes to the explosion in particular near the polar region. Due to the mass ejection associated with the jet, an outflow is driven from the inner region of the torus, and then, some of the ejected matter has relatively low values (0.3--0.4) of the electron fraction, $Y_\mathrm{e}$, because the inner region of the torus has high density, i.e., the electron degeneracy is high, for this model (note that this may not be the case for other models). The star entirely explodes at $t \sim 30$\,s for B11.1.8h. 

For this model, a torus component with its rest-mass lower than $\rho_\mathrm{cut}$ remains even at the stellar explosion. This is an artifact for the choice of the high cutoff density to the dynamo parameter, below which density turbulence of the disk/torus is inactive. For the lower cutoff density, the mass of the left-over torus is much smaller and the ejecta mass is larger (cf.~Table~\ref{tab2}; e.g., compare the results of models B12.1.8l and B12.1.8h). For the low cutoff density, the initial mass minus the sum of the final black hole mass and ejecta mass is less than $1M_\odot$ for many models, and hence, we consider that the effect for the artificial choice of the cutoff density is likely to be minor for $\rho_\mathrm{cut}=10^6\,\mathrm{g/cm^3}$. 

For the present choices of the dynamo parameters, the features described above are also observed for many models. However, for a few runs (B12.1.7l, B12.1.7l-H,  and B12.3.8l-H), the jet power is weak or jet is not seen until the termination of the simulation, although the stellar explosion with the explosion energy larger than $10^{51}$\,erg is universally achieved by the torus activity  (cf.~Table~\ref{tab2}). For these runs, the poloidal magnetic field that penetrates the black hole is not established well. Our interpretation for this is as follows: First, the magnetic field enhancement is achieved by the turbulent process, i.e., a stochastic process. Even if a seed poloidal magnetic field penetrates the black hole, it could pair-annihilate due to the infall of a magnetic flux with a different polarity from the torus. This process could prevent the formation of the swift formation of a large-scale poloidal field. Note that the poloidal magnetic field and resultant jet power become weak if the global poloidal magnetic field is established in a late stage; see Eq.~\eqref{eq1}. This is the case for models B12.1.7l and B12.1.7l-H. For these models, the dissipation efficiency of the magnetic field is high, and this may affect the jet launch timing. Second, a strong poloidal field should be developed when the density and gas pressure of the torus are high enough, i.e., the magnetic field strength is high enough (cf.~Eq.~\eqref{eq8}). For the case that the value of $\alpha_\mathrm{d}$ is large (and $\sigma_\mathrm{c}$ is not very low), the dynamo can be efficiently activated, resulting in a quick increase of the thermal energy and expansion of the torus. In such a situation, the mass ejection from the torus can be driven in a relatively early stage, while the accretion of the magnetic flux with a strong magnetic field onto the black hole is prevented. Then, a stellar explosion with large energy from the torus engine is possible, but the jet launch can be inactive. This is seen for model B12.3.8l-H. As a consequence, this model is similar to the viscous hydrodynamics model in which only a high-energy stellar explosion (no jet launch) is the result (cf.~Table~\ref{tab2}). 

By contrast, for some models like B12.1.8l and B12.3.8l, the energies of both the stellar explosion and the jet are high. For these models, a large-scale poloidal magnetic field that penetrates the black hole is established in a relatively early stage, leading to a high field strength (see Eq.~(\ref{eq1})). For these early jet-launch models, the stellar explosion also sets in at earlier time with large explosion energy (cf.~Table~\ref{tab2}). All these results illustrate that for strong jet launching, a good combination of the dynamo parameters, i.e., a suitable activity for the dynamo, is required. 

Stellar explosion was also found in our viscous hydrodynamics simulation~\cite{Fujibayashi:2023oyt} and in our ideal magnetohydrodynamics simulation~\cite{Shibata:2023tho}. For the viscous hydrodynamics case, the explosion mechanism is qualitatively similar to that found in this paper; the viscous heating in the torus in the vicinity of the black hole can be the primary engine for the explosion. On the other hand, the explosion mechanism in our previous ideal magnetohydrodynamics simulation is different from those of the viscous hydrodynamics and present magnetohydrodynamics ones because, in the previous ideal magnetohydrodynamics, the turbulence in the torus was not excited. For that case, the magnetocentrifugal force resulting from the black-hole spin or the differential rotation in the torus together with a high field strength is the primary engine. Since relatively strong poloidal magnetic fields were initially {\em assumed} in Ref.~\cite{Shibata:2023tho}, such a mechanism can work for inducing the stellar explosion (i.e., in a sense, the result is determined by the assumption/initial setting). In the present study in which the initial poloidal magnetic field is not extremely strong at the formation of the torus, this mechanism is subdominant. 

\subsection{Evolution of the black hole}

Figure~\ref{fig3} shows the evolution of the mass and dimensionless spin of the black holes for models B11.1.8h, B12.1.8h, B12.1.8l, B12.3.8l, B12.1.7l, and B12.1.9l (left) and for models B12.1.8l, B12.3.8l, B12.1.7l, and their high-resolution runs (right). For comparison, we also plot the results of a new viscous hydrodynamics simulation in which we also find the stellar explosion (but no jet launch; see Ref.~\cite{Fujibayashi:2023oyt} for the results in a similar setting). As we already mentioned, these quantities increase monotonically and steadily with time in an early stage of the evolution (for $t \alt 10$\,s). After the development of a massive torus, their increase rates decrease, and after the onset of the stellar explosion, the values of these quantities approximately relax to constants, as $M_\mathrm{BH}\approx 20$--$22M_\odot$ and $\chi\approx 0.78$--0.84. 
The feature is also quite similar to that in our previous paper~\cite{Shibata:2023tho}, although the stellar explosion mechanism in that paper, which was caused solely by the magnetocentrifugal effects, is different from that in the present paper; from the evolution curves of the black hole, the mechanism of the explosion cannot be identified. 

We here stress that in the present work, the poloidal magnetic-field strength at the black-hole horizon that accounts for the jet launch is $\sim 10^{14}$\,G, and as a result, the spin-down timescale associated with the Blandford-Znajek mechanism is much longer than 100\,s in contrast to those in the strongly magnetized models of Ref.~\cite{Shibata:2023tho}. In the present result, the spin-up by the mass accretion always dominates over the spin-down by the Blandford-Znajek mechanism. 

By a close look at the final values of $M_\mathrm{BH}$ and $\chi$, we find that for models in which earlier explosion is induced, the resulting mass and dimensionless spin of the black holes are smaller. This makes it possible to eject larger mass and to synthesize more $^{56}$Ni (cf. Tables~\ref{tab2} and \ref{tab3}). 

For the viscous hydrodynamics result, both the final mass and dimensionless spin of the black hole are smaller than those of many of the magnetohydrodynamics results. The reason for this is that in our viscous hydrodynamics, the viscosity turns on from $t=0$, and hence, soon after the development of the massive disk/torus, the viscous activity is enhanced, leading to an earlier stellar explosion. On the other hand, in the present magnetohydrodynamics simulations, it takes more time until the magnetic field strength is sufficiently enhanced and a turbulent state is established. Thus, the stellar explosion and jet launch are induced in later stages. 

\begin{figure*}[t]
\includegraphics[width=0.495\textwidth]{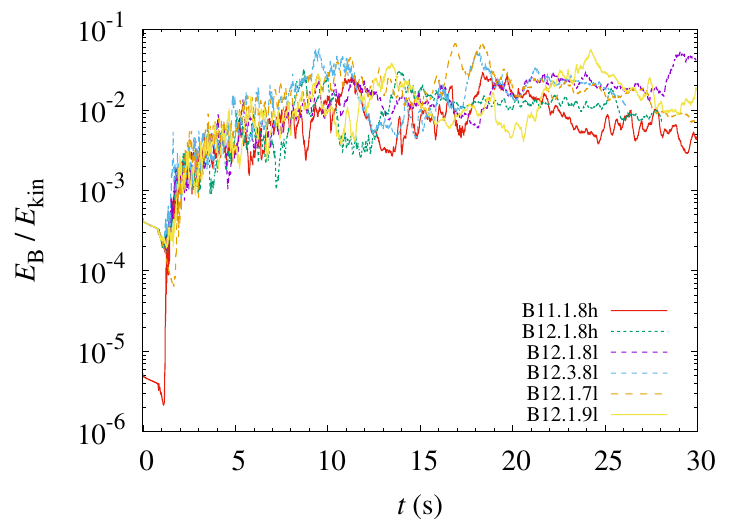}
\includegraphics[width=0.495\textwidth]{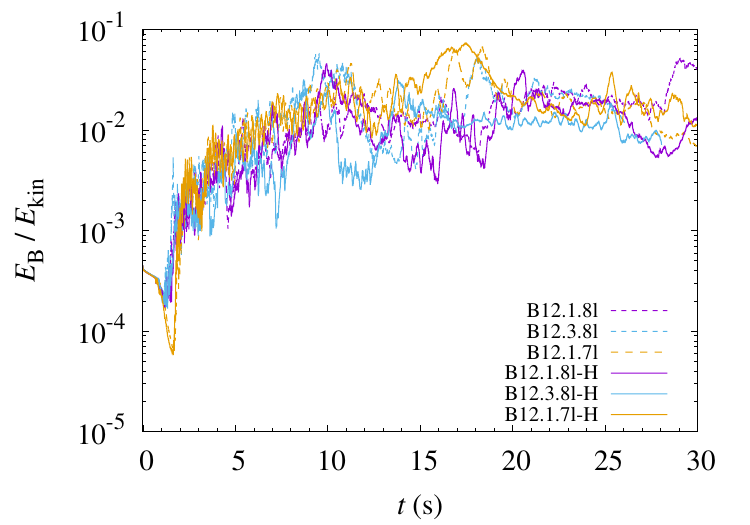}
\caption{The same as Fig.~\ref{fig4} but for the ratio of the electromagnetic energy to the kinetic energy of the matter. 
\label{fig5}}
\end{figure*}

\subsection{Evolution of the electromagnetic energy}

We confirm that the strength of the poloidal magnetic field that penetrates the black hole becomes higher for the earlier jet launch model as Eq.~(\ref{eq1}) indicates. This is reflected in the high Blandford-Znajek Poynting flux for models such as B12.1.8l, B12.1.8l-H, and B12.3.8l (see the next subsection). 

By contrast, the evolution curves of the electromagnetic energy and the ratio of the electromagnetic energy to the kinetic energy do not show noticeable differences among the models. Figure~\ref{fig4} displays the evolution of the electromagnetic energy with time. Before the formation of a disk at $t \sim 1.5$\,s, it never grows but rather slightly decreases with time during the matter infall due to the toroidal nature of the initial magnetic-field profile. 
Once the disk (and subsequently a torus) is developed, the dynamo effect enhances the magnetic-field strength significantly until the saturation is reached. The saturation time is $\sim 10$\,s irrespective of the models. After the saturation, the electromagnetic energy is kept in average of order $10^{50}$\,erg, and at the onset of the stellar explosion, it starts decreasing with time because the magnetic-field strength becomes weaker as the explosion progresses. 

The ratio of the electromagnetic energy to the kinetic energy, displayed in Fig.~\ref{fig5}, steeply increases at $t \sim 1.5$\,s and subsequently until $t \sim 10$\,s, it gradually increases to $\sim 10^{-2}$. Our interpretation of this slow increase is due to the fact that before $t\sim 10$\,s, the kinetic energy of the infalling matter (not the rotational kinetic energy of the torus) contributes primarily to the total kinetic energy. However, for $t \agt 10$\,s, this ratio relaxes to $\sim 10^{-2}$, indicating that the torus is massive, turbulent, and in an equi-partition state, which seems to be prerequisite for the subsequent jet launch and stellar explosion from the torus. 


We here touch on the convergence of the numerical results. In the right panels of Figs.~\ref{fig3}--\ref{fig5}, we compare the numerical results with two different grid resolutions. These figures indicate that numerical convergence is well achieved before the turbulence in the torus is excited. On the other hand, after the development of the turbulence, the convergence is fair. This is reasonable because the stochastic motion is dominant in the stage in which the turbulence is excited. Thus, the results in this paper for the explosion and jet are qualitatively and semi-quantitatively reliable. However we have to keep in mind that quantitatively the numerical results in the explosion energy, ejecta mass, and Poynting luminosity, which will be shown in the next subsection, would have an uncertainty of a factor of a few. This is in particular the case for the quantity associated with the jet driven by the Blandford-Znajek mechanism because the magnetosphere around the spin axis of the black hole is established after the turbulence is significantly excited in the torus. For example, for model B12.3.8l we find a jet launch whereas for model B12.3.8l-H we do not find, although for both models we find the stellar explosion with high explosion energy: This reflects the fact that for B12.3.8l-H, the electromagnetic power from the torus is predominantly used for the stellar explosion toward the direction different from the spin axis. 

\subsection{Explosion energy, ejecta mass, and Poynting luminosity}
\label{secIIIB}

\begin{table}[t]
\centering
\caption{Key results. From left to right, approximate time at which the explosion sets in, ejecta mass, explosion energy, Poynting luminosity integrated over simulation time, and total energy emitted by neutrinos, respectively.
\label{tab2}
}
\begin{tabular}{lcccccc}
\hline
Model~~~ & ~~$t_\mathrm{exp}$~~ & $M_\mathrm{ej}$ & $E_\mathrm{exp}$   & $E_\mathrm{BZ}$ & $E_\nu$ \\
     & (s) &  ($M_\odot$)    &  ($\SI{e51}{erg}$) &  ($\SI{e51}{erg}$) & 
      ($\SI{e53}{erg}$) \\
\hline
B11.1.8h   & 17.0 &   1.56  &    1.68  &    1.20 & 3.96 \\ 
B12.1.8h   & 14.3 &   1.58  &    1.33  &    0.87 & 3.16 \\ 
B12.1.8l   & 11.7 &   3.20  &    5.40  &    1.20 & 2.73 \\ 
B12.3.8l   &  7.6 &   4.90  &   11.6   &    1.59 & 1.86\\ 
B12.1.7l   & 17.4 &   1.77  &    1.67  &    0.22 & 3.18 \\ 
B12.1.9l   & 20.7 &   2.86  &    2.97  &    0.51 & 3.16 \\ 
\hline
B12.1.8l-H   &  10.1 & 2.40 & 3.98  & 0.83 & 3.53   \\ 
B12.3.8l-H   &  7.1  & 3.55 & 4.65  & 0.93 & 1.85 \\ 
B12.1.7l-H   &  15.0 & 1.95 & 1.92  & 0.20 & 3.12 \\ 
\hline
viscous & 5.2 & 4.80 & 8.31 & -- & 0.34 \\ \hline
\end{tabular}\\
\end{table}


\begin{figure*}[t]
\includegraphics[width=0.49\textwidth]{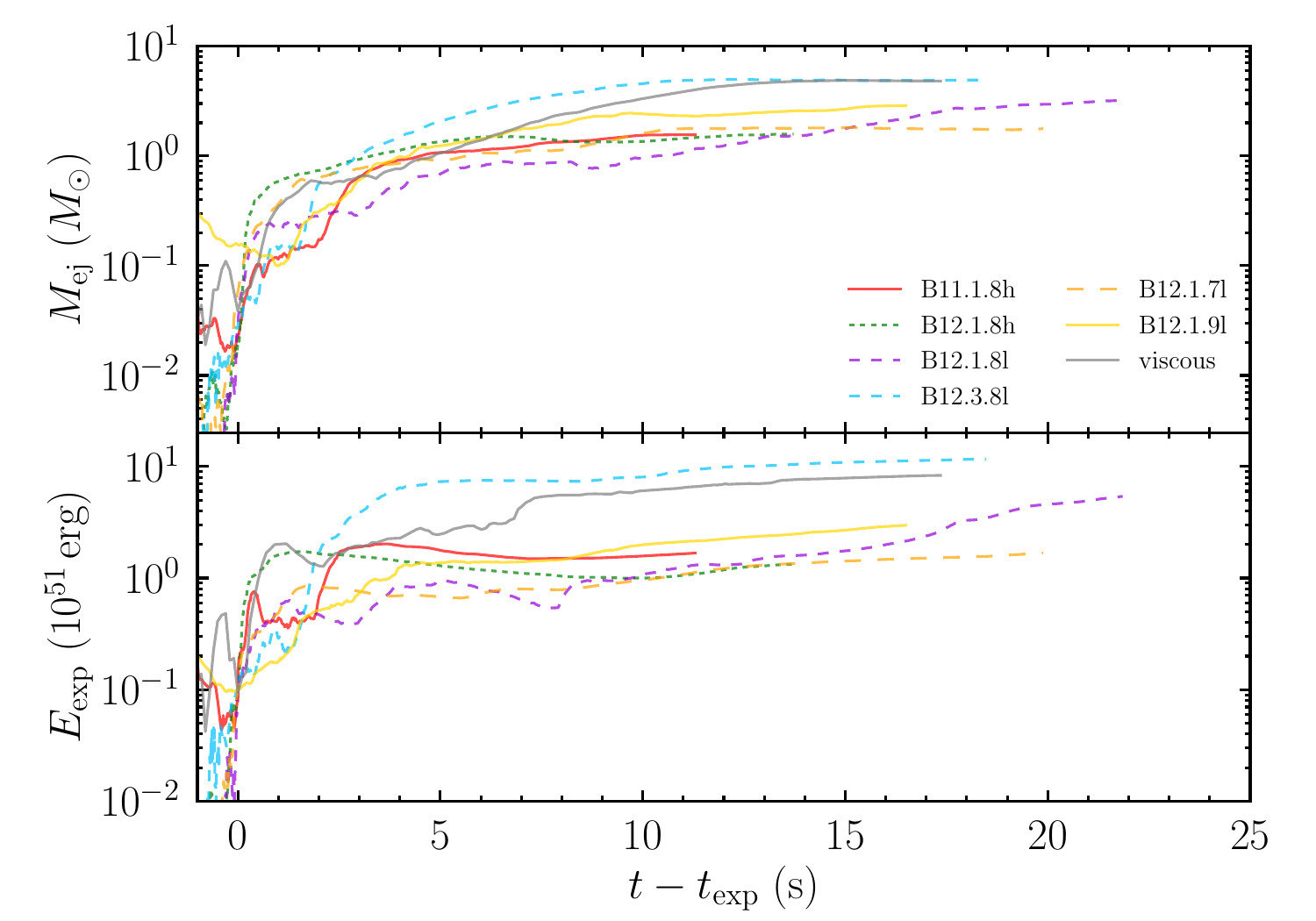}
\includegraphics[width=0.49\textwidth]{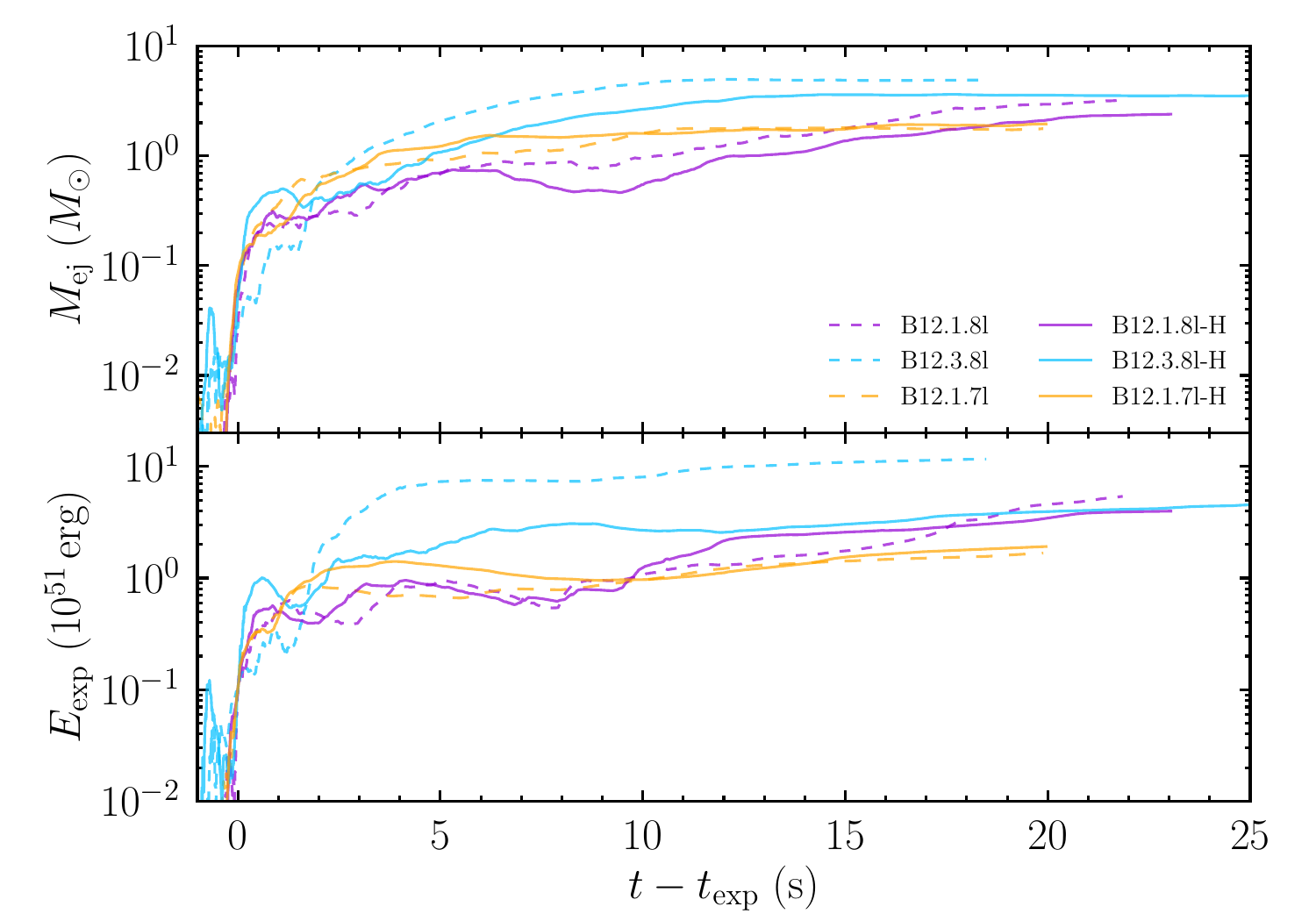}
\caption{Left: Time evolution of the ejecta mass (top panel) and explosion energy (bottom panel) for all the standard-resolution models as well as for the viscous model. The time origin for each model is chosen to be the time of the onset of the explosion, i.e., $t_\mathrm{exp}$. Right: comparison of the results in two different grid resolutions. 
}
\label{fig:MejEexp}
\end{figure*}

\begin{figure*}
\includegraphics[width=0.95\textwidth]{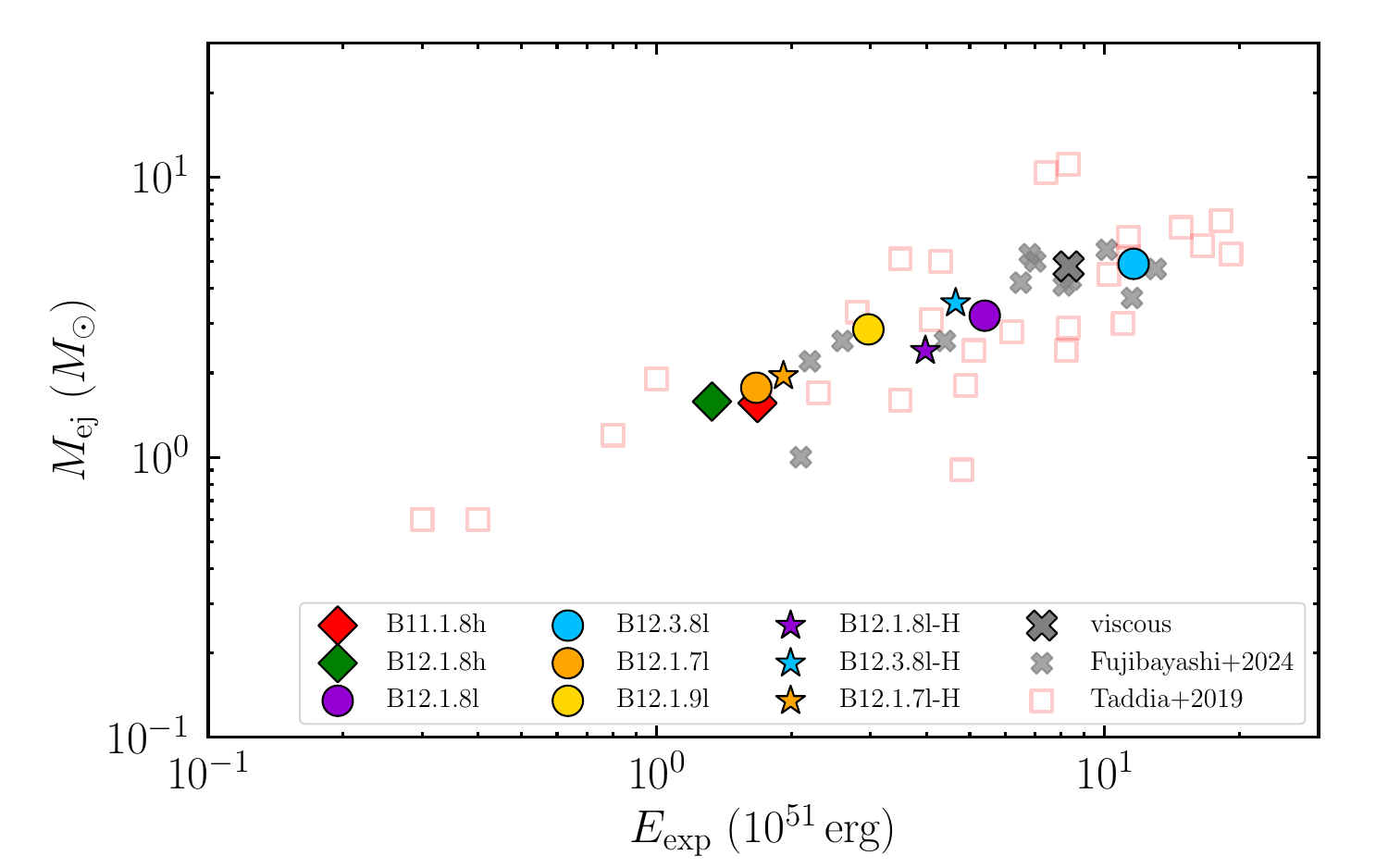}
\caption{Correlation between the ejecta mass and explosion energy (filled markers). The open markers denote the inferred values from the observations of type Ic-BL SNe \citep{Taddia2019jan}. The grey crosses denote the results of viscous hydrodynamics simulations of Ref.~\citep{Fujibayashi2024jan}. The diamonds and circles denote the models with high and low cut-off density $\rho_\mathrm{cut}$, respectively. The stars denote the results of the runs with a higher grid resolution. Note that in Ref.~\citep{Fujibayashi2024jan}, progenitor's angular momentum was varied for a wide variety and the diversity of $M_\mathrm{ej}$ and $E_\mathrm{exp}$ reflects this variation. 
}
\label{fig:ejecta-Eene}
\end{figure*}


\begin{figure}
\includegraphics[width=0.495\textwidth]{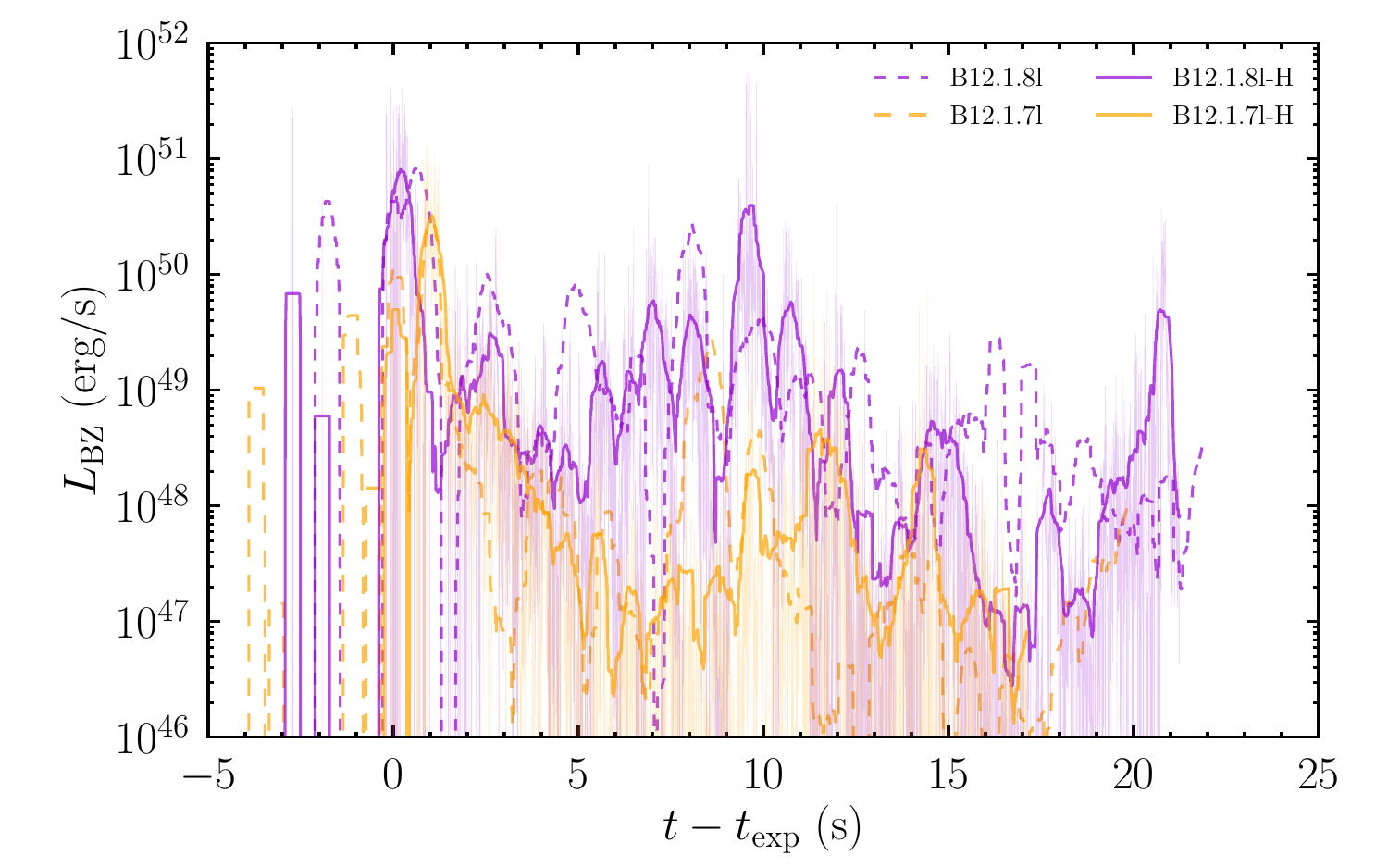}
\includegraphics[width=0.495\textwidth]{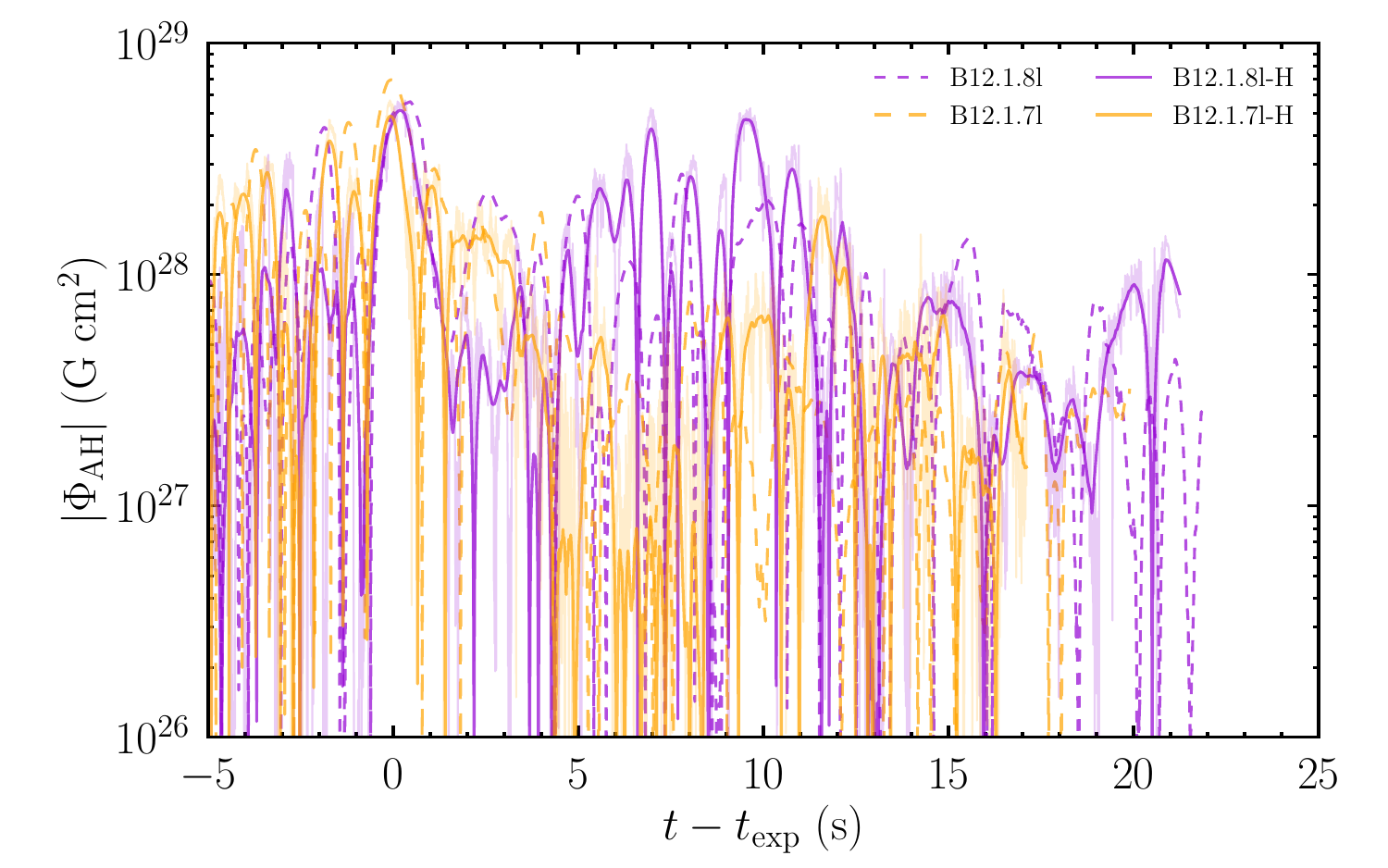}
\includegraphics[width=0.495\textwidth]{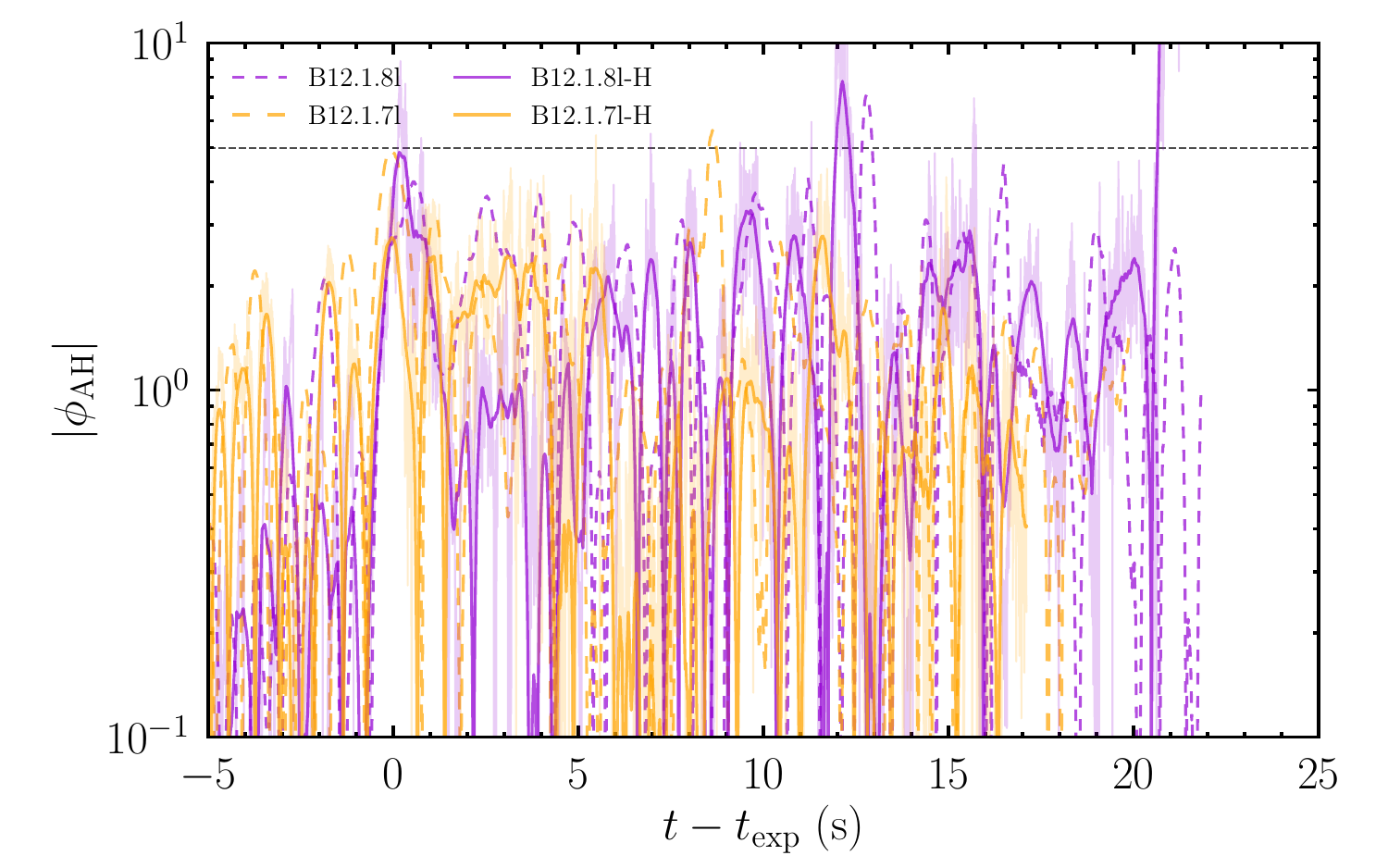}
\caption{Time evolution of the Poynting luminosity (top), the magnetic flux at the apparent horizon (middle), and MADness parameter (bottom) for models B12.1.8l (high-luminosity model), B12.1.7l (low-luminosity model), and their higher-resolution runs. For each panel, the thick curves denote their moving average with a time interval of 0.2~s. The thin curves denote the original data of higher-resolution runs sampled with a time interval of 0.5~ms.  In the bottom panel, the horizontal dotted line denotes $|\phi_\mathrm{AH}|=5$.
}
\label{fig:BZ}
\end{figure}

The approximate onset time of the explosion $t_\mathrm{exp}$, total ejecta mass $M_\mathrm{ej}$, explosion energy $E_\mathrm{exp}$, and Poynting energy associated with the Blandford-Znajek mechanism $E_\mathrm{BZ}$ are summarized in Table~\ref{tab2}. Note that in the explosion energy, the electromagnetic energy is included but it is always subdominant. Figure~\ref{fig:MejEexp} shows the evolution of the ejecta mass, $M_\mathrm{ej}$, and explosion energy, $E_\mathrm{exp}$,  as functions of $t-t_\mathrm{exp}$ for all the models simulated in this paper. It is found that $M_\mathrm{ej}$ and $E_\mathrm{exp}$ increase gradually with time after the sudden increase at the onset of the explosion irrespective of the models. The engine of this increase is the long-term magnetohydrodynamical activity of the torus. 

Broadly speaking, for earlier explosion models, i.e., B12.1.8l,  B12.1.8l-H, and B12.3.8l, $M_\mathrm{ej}$, $E_\mathrm{exp}$, and $E_\mathrm{BZ}$ are all relatively large (this is also the case for the viscous hydrodynamics model for which the explosion always sets in earlier), while for late-explosion models such as B12.1.8h and B12.1.7l, they tend to become relatively low. This is reasonable because for the late-explosion models, the matter located outside the black hole is smaller at the onset of the explosion, and the poloidal magnetic field established at the jet launch is weaker (see Eq.~(\ref{eq1})). However, the explosion energy is always comparable to or larger than ordinary SNe, i.e., $\agt 10^{51}$\,erg, for all our models, as in our viscous hydrodynamics models~\cite{Fujibayashi:2023oyt}. In particular for models B12.1.8l, B12.1.8l-H, B12.3.8l, and B12.3.8l-H, for which the explosion sets in relatively earlier, the explosion energy is much higher than those of the ordinary SNe and comparable to the energetic SNe such as broad-lined type Ic (type Ic-BL) SNe~\cite{Cano2017a} (see below). For these models (except for B12.3.8l-H for which no jet is launched), the total energy of the Poynting flux, $E_\mathrm{BZ}$, is also comparable to the energy required for long GRBs~\cite{2008ApJ...675..528L}, and hence, these are good models for GRB-SN events. It should be also mentioned that for other models, $E_\mathrm{BZ}$ is still of order $10^{50}$\,erg, which can account for some of long GRBs. 


The explosion energy, $E_\mathrm{exp}$, is always larger than $E_\mathrm{BZ}$. In particular, for high-explosion energy models, $E_\mathrm{exp}$ is by a factor of several larger than $E_\mathrm{BZ}$. This suggests that the stellar explosion energy would be much larger than the energy of GRBs in our present scenario. This result is consistent with the finding of Ref.~\cite{Eisenberg2022nov}.

The filled markers of Fig.~\ref{fig:ejecta-Eene} show the correlation between the ejecta mass and explosion energy. The open markers denote the inferred values from the observations of type Ic-BL SNe \citep{Taddia2019jan}. The grey crosses denote the results of viscous hydrodynamics simulations~\citep{Fujibayashi2024jan}. The stars denote the results of the runs with a higher grid resolution. It is found that as in the viscous hydrodynamics results of Refs.~\cite{Fujibayashi2024jan} (see also Refs.~\cite{Just2022aug,Dean_2024,Menegazzi2024arXiv}), the present results, which are obtained purely from the results of the numerical simulations with no fine tuning, are in a good agreement with the observational results. It is worth stressing that the correlation between the explosion energy and ejecta mass is quantitatively reproduced. This suggests that the present collapsar scenario can be a good candidate for interpreting the observational data of the energetic SNe. 

We here note that the diversity of $M_\mathrm{ej}$ and $E_\mathrm{exp}$ in viscous hydrodynamics simulations of Ref.~\cite{Fujibayashi2024jan} arose from the variation of the progenitor's angular momentum; for the higher value of the progenitor's angular momentum, these values are larger. For the present work, on the other hand, the diversity of $M_\mathrm{ej}$ and $E_\mathrm{exp}$ arises from the variety of the dynamo parameters, i.e., from the degree of the dynamo activity. If the progenitor model is in addition varied for a wide variety, the diversity of $M_\mathrm{ej}$ and $E_\mathrm{exp}$ would be even wider. 

Another point to be noted is that the ejecta mass in our result is always smaller than $5M_\odot$, and hence, our models cannot explain the observational data with ejecta mass of $\agt 5M_\odot$. In our present choice of the progenitor model, the total baryon mass at the initial condition is $\alt 10M_\odot$, and thus, it is in principle impossible to reproduce the observational data with $M_\mathrm{ej}\geq 10M_\odot$. To reproduce the data with a high value of the ejecta mass, thus, different progenitor models are necessary.  

In our resistive magnetohydrodynamics simulations, the strength of the poloidal magnetic fields that penetrate the black hole decreases gradually over time because the magnetic fluxes with different polarities are ejected from the torus and weaken the magnetic-field strength in the magnetosphere via reconnection. The top panel of Fig.~\ref{fig:BZ} shows the Poynting flux extracted from the black hole as a function of $t-t_\mathrm{exp}$ for selected models with high luminosity (B12.1.8l abd B12.1.8l-H) and low luminosity (B12.1.7l and B12.1.7l-H). We note that for other models the results have quantitatively similar dependence on $E_\mathrm{BZ}$. Soon after the jet launch, the peak is reached with $L_\mathrm{BZ} \sim 10^{51}\,$erg/s for B12.1.8l and B12.1.8l-H and with $\sim 10^{50.5}$\,erg/s for B12.1.7l and B12.1.7l-H. For $\sim 10$\,s, the peak values of $L_\mathrm{BZ}$ are kept to be $\agt 10^{50}$\,erg/s for B12.1.8l and B12.1.8l-H, but subsequently, it decreases with time. This is due to the decrease of the poloidal magnetic-field strength via the reconnection. For B12.1.7l and B12.1.7l-H, the values of $L_\mathrm{BZ}$ drops more quickly after the first peak.

The middle panel of Fig.~\ref{fig:BZ} shows the absolute value of the poloidal magnetic flux that penetrates the black hole, $\Phi_\mathrm{AH}$. It is found that this quantity decreases gradually over time for $t \geq t_\mathrm{exp}$ as well. We find that the magnetic fluxes for models B12.1.8l and B12.1.8l-H are higher than those for models B12.1.7l and B12.1.7l-H. The reason for this is that the magnetosphere along the spin axis is developed earlier for models B12.1.8l and B12.1.8l-H (see Table~\ref{tab2} and Eq.~(\ref{eq1})).  
We find an oscillation for $\Phi_\mathrm{AH}$ with time irrespective of the models. This is caused by the magnetic polarity changes with time due to the dynamo action of the accretion torus and resulting reconnection of the magnetic field in the magnetosphere. 
This mechanism suppresses the over-extraction of the rotational kinetic energy of the black holes by the Blandford-Znajek mechanism, leading to a reasonable value of the total extracted energy of order $10^{51}$\,ergs (see Table~\ref{tab2}). We note that the result obtained in this paper is quite different from that in our previous ideal-magnetohydrodynamics work~\cite{Shibata:2023tho} for which high luminosity is preserved for much longer timescales, because in the previous axisymmetric and ideal magnetohydrodynamics work no dynamo effect was taken into account. 

The bottom panel of Fig.~\ref{fig:BZ} plots the evolution of the MADness parameter defined by
\beq
\phi_\mathrm{AH}:={\Phi_\mathrm{AH} \over \sqrt{4\pi G^2 c^{-3} \dot M_\mathrm{BH,*} M_\mathrm{BH}^2}},
\eeq
where $\dot M_\mathrm{BH,*}$ is the rest-mass accretion rate onto the black hole. It is found that this value is at most $\sim 5$, which is much smaller than the critical value for establishing the fully magnetically arrested disk (MAD) state $\sim 50$~\cite{Igumenshchev:2003rt, Tchekhovskoy:2011zx}. The reason for this is that the MAD state is established only locally (near the spin axis of the black hole) in our present models. In our models the condition for launching a jet with certain strength is $\phi_\mathrm{AH}\agt 5$; for model B12.1.7l, the jet is weak. This condition is quantitatively similar to that found in Ref.~\cite{Shibata:2023tho} as well as that found in black hole-neutron star merger simulations~\cite{Hayashi:2021oxy, Hayashi:2022cdq}. 

As found from the middle and bottom panels of Fig.~\ref{fig:BZ}, in the present simulations, the steady poloidal magnetic-field structure cannot be developed. The reason for this is that in the presence of the dynamo action, the polarity of the magnetic fields change in the torus in a short timescale quasi-periodically, and hence, the poloidal magnetic field developed in the magnetosphere is soon damaged by a magnetic flux with a different polarity coming from the torus~\cite{2008ApJ...678.1180B}. No establishment of a quasi-steady magnetosphere with aligned poloidal magnetic fields in the polar region may be due to our very simple modeling of the dynamo process, because an aligned and quasi-steady poloidal magnetic field is often established for a longer timescale in other ideal magnetohydrodynamics works (e.g., Refs.~\cite{Christie2019dec,Hayashi:2021oxy, Hayashi:2024jwt}). A simulation with more sophisticated dynamo modeling is an issue left for the future.

\subsection{Nucleosynthesis, Nickel mass, and $r$-process elements} \label{sec4E}

\begin{figure*}
\includegraphics[width=0.495\textwidth]{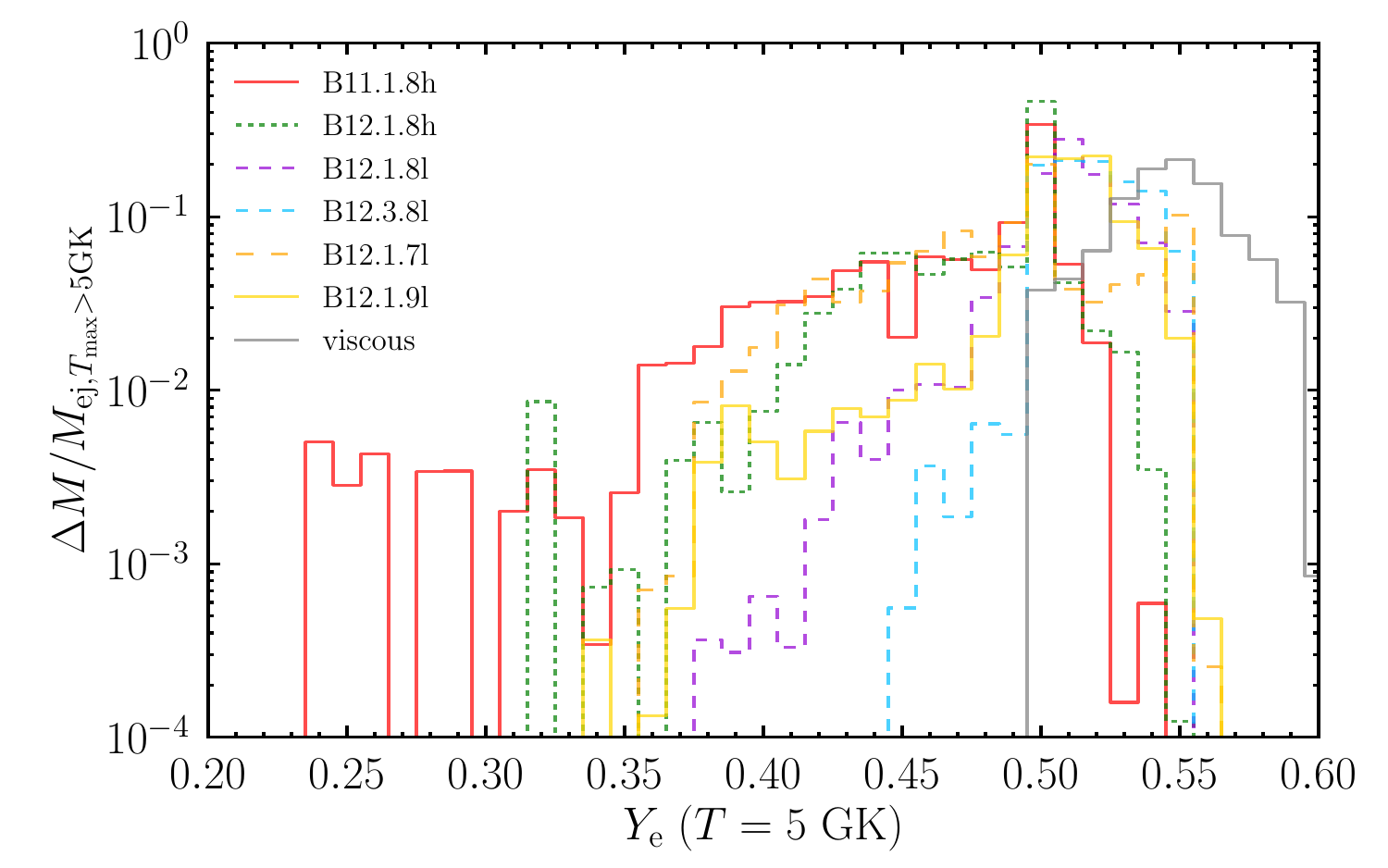}
\includegraphics[width=0.495\textwidth]{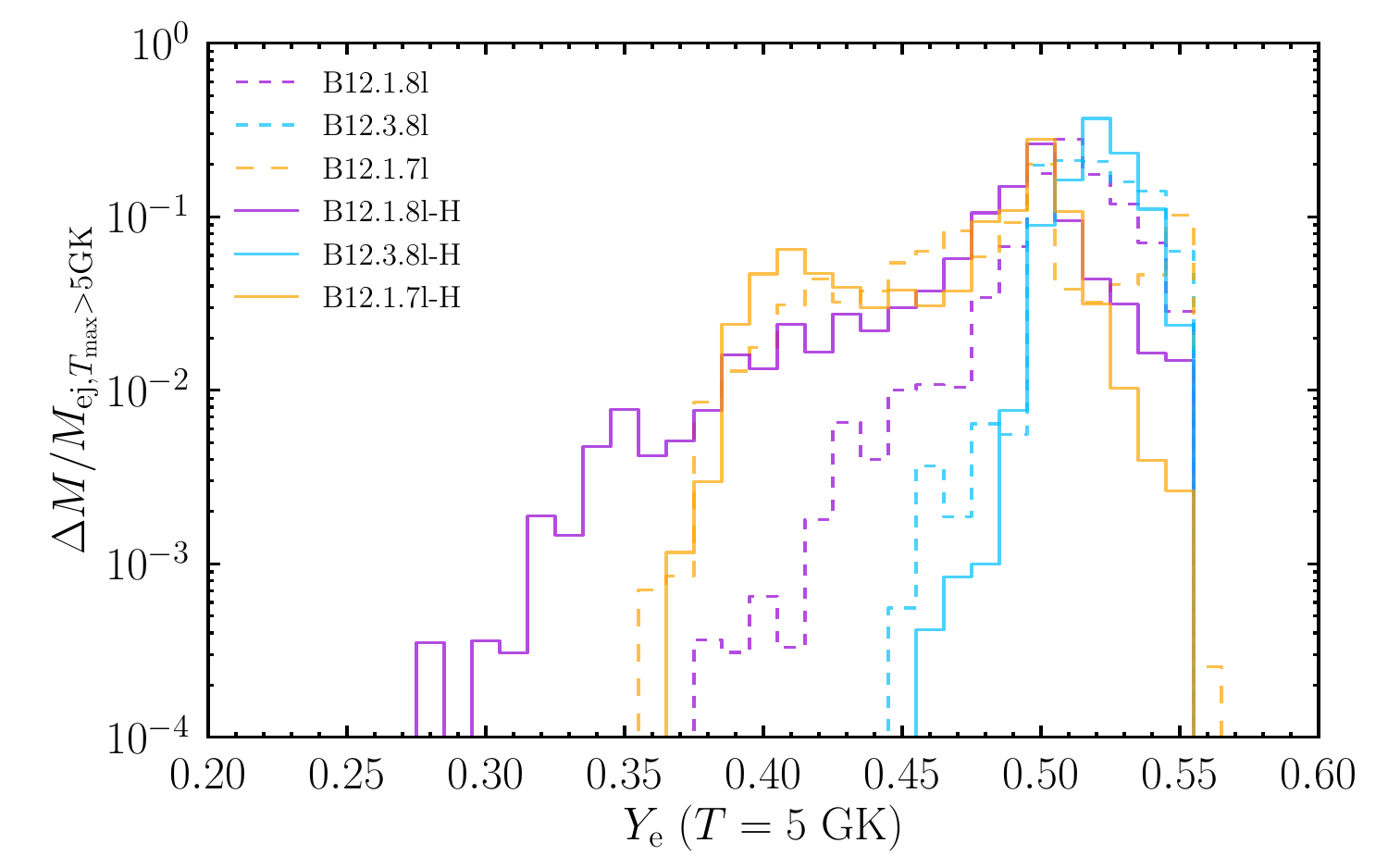}\\
\includegraphics[width=0.495\textwidth]{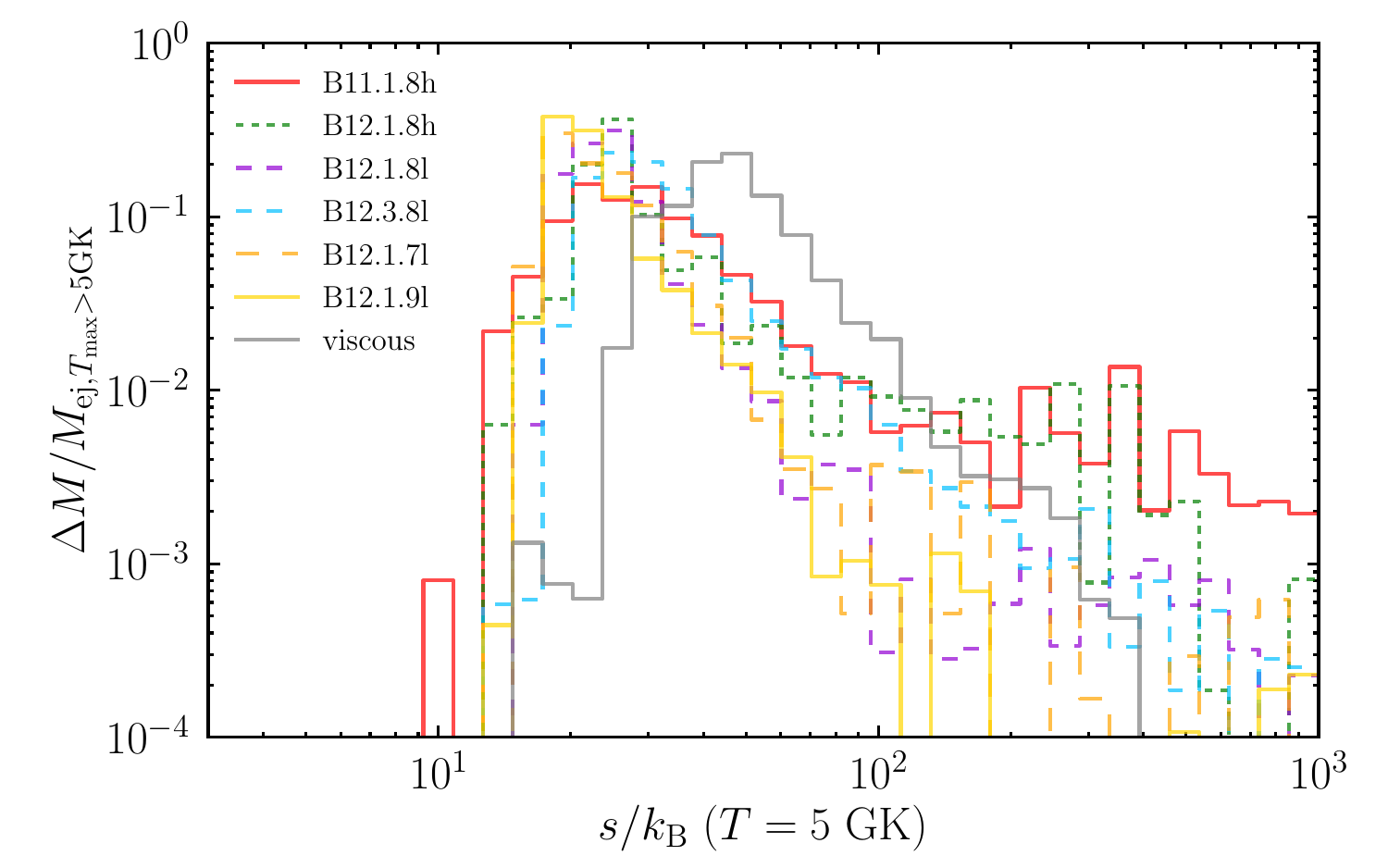}
\includegraphics[width=0.495\textwidth]{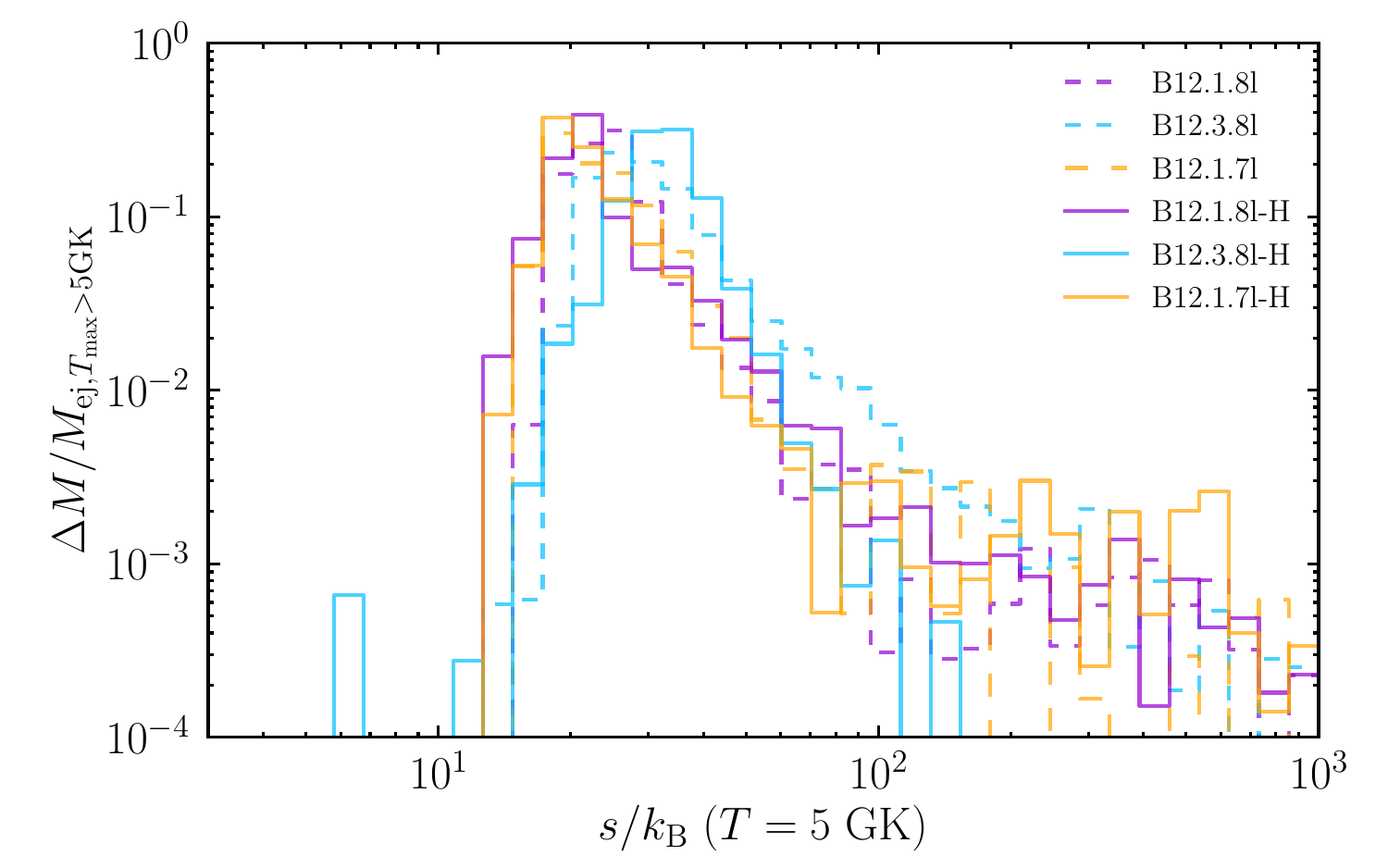}\\
\includegraphics[width=0.495\textwidth]{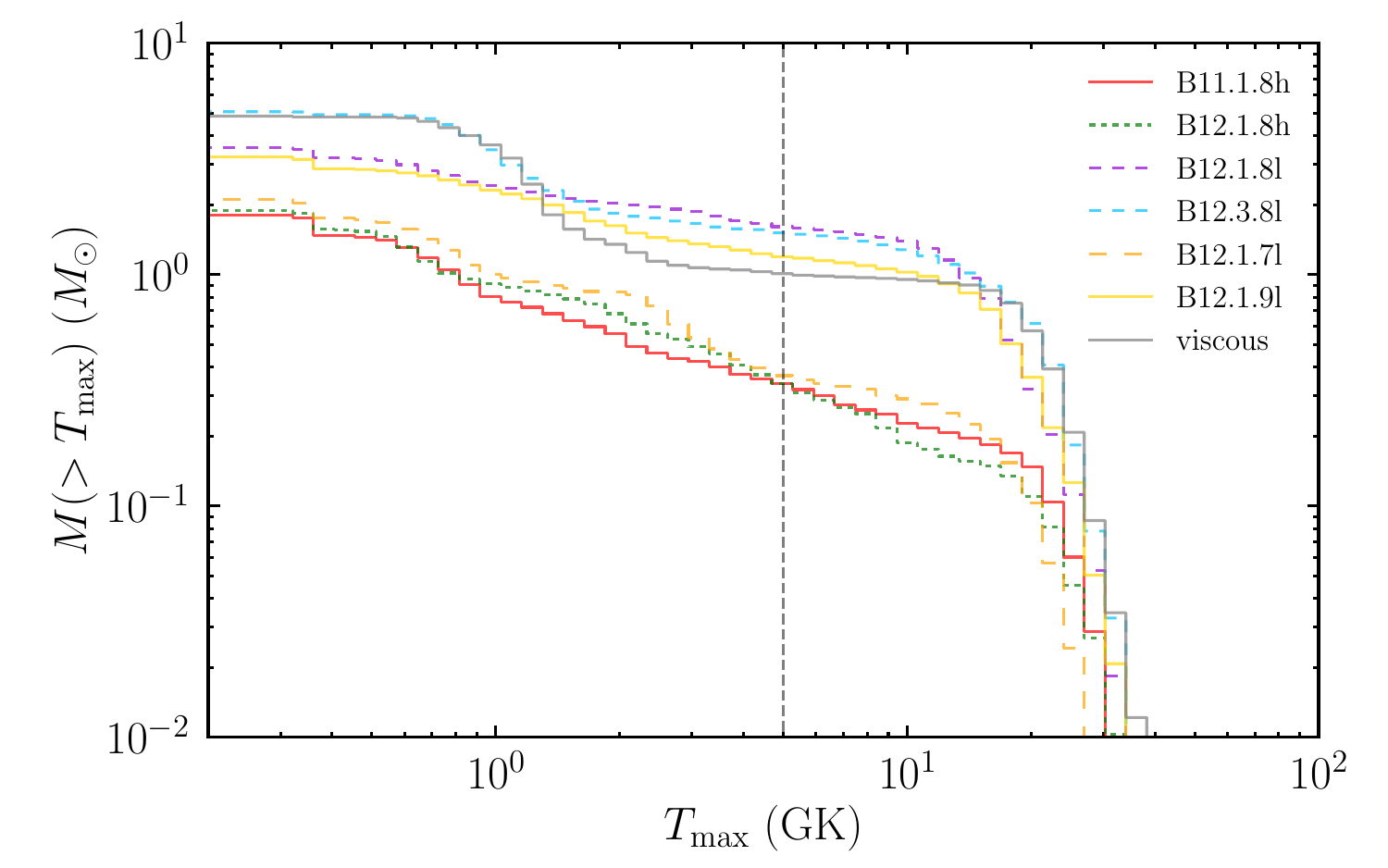}
\includegraphics[width=0.495\textwidth]{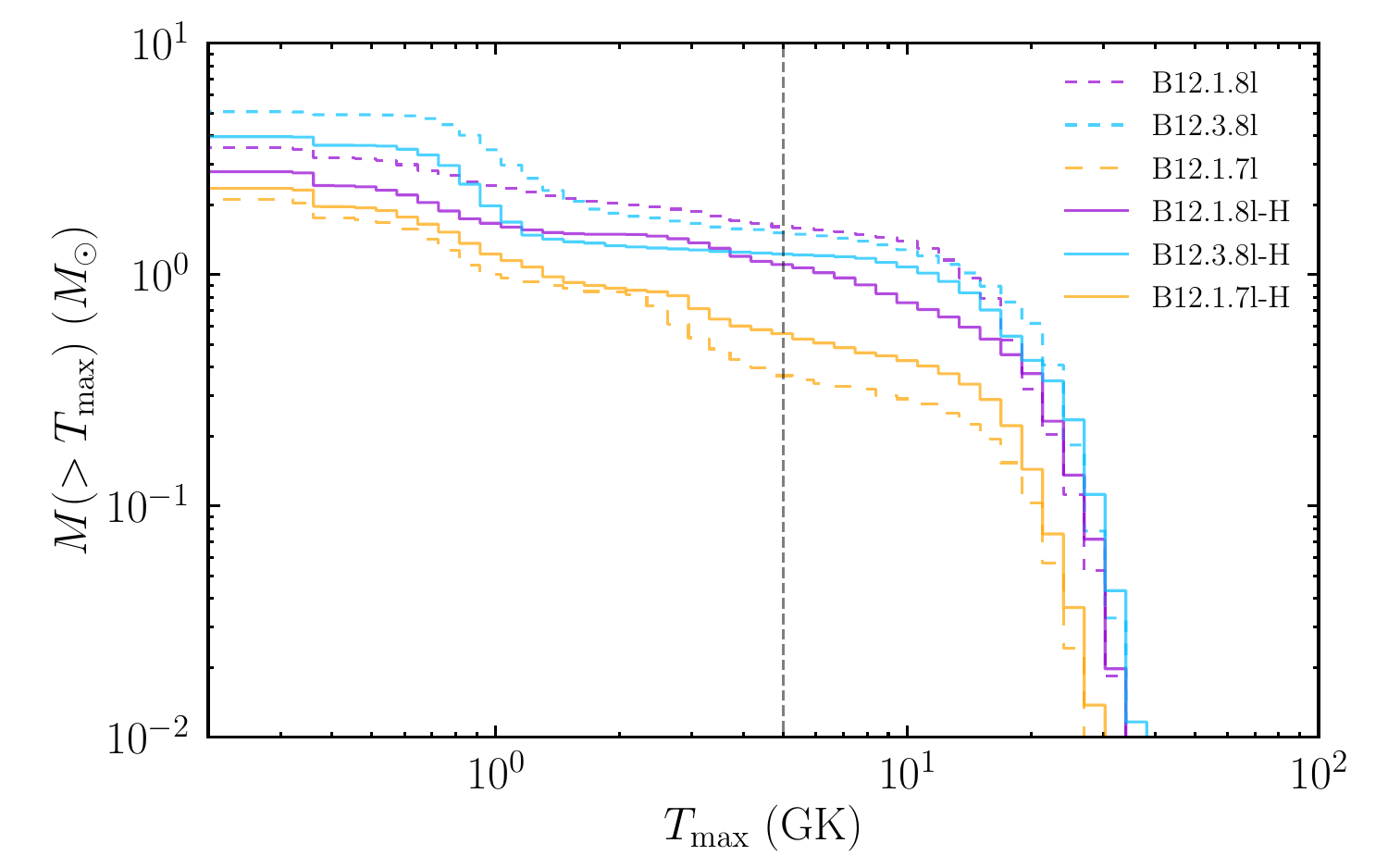}
\caption{Mass histogram of electron fraction (top panels), entropy per baryon (middle panels), and cumulative distribution of the maximum temperature that the ejecta ever experienced (bottom panels). The left panels show the results for all the standard-resolution runs, and the right panels compare the results with different grid resolutions.  
}
\label{fig:cumulative}
\end{figure*}

\begin{figure*}
\includegraphics[width=0.495\textwidth]{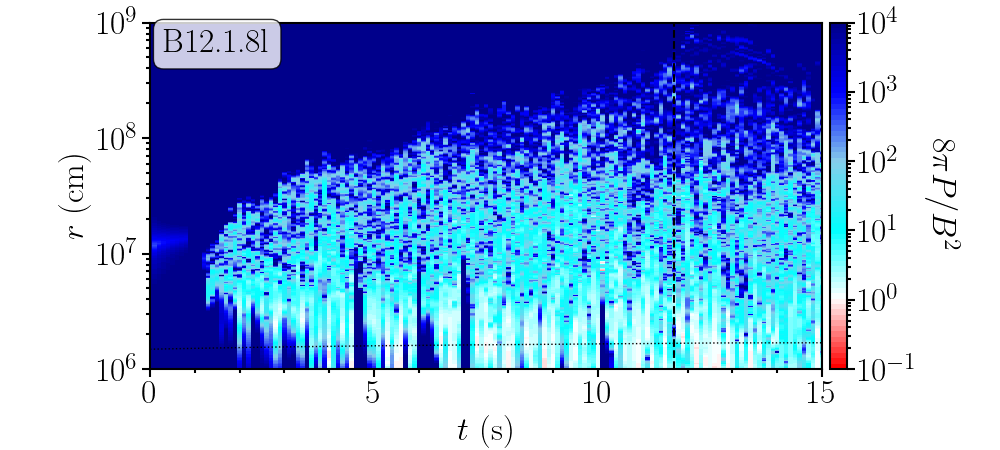}
\includegraphics[width=0.495\textwidth]{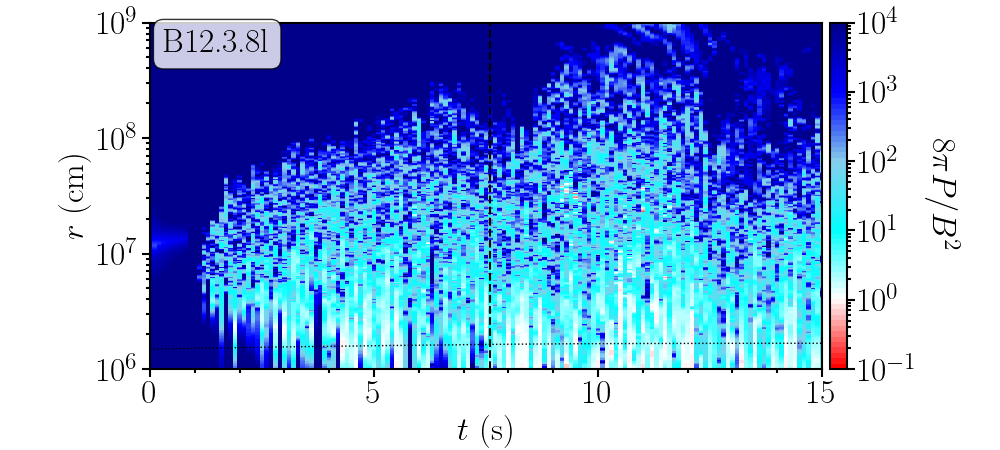}
\includegraphics[width=0.495\textwidth]{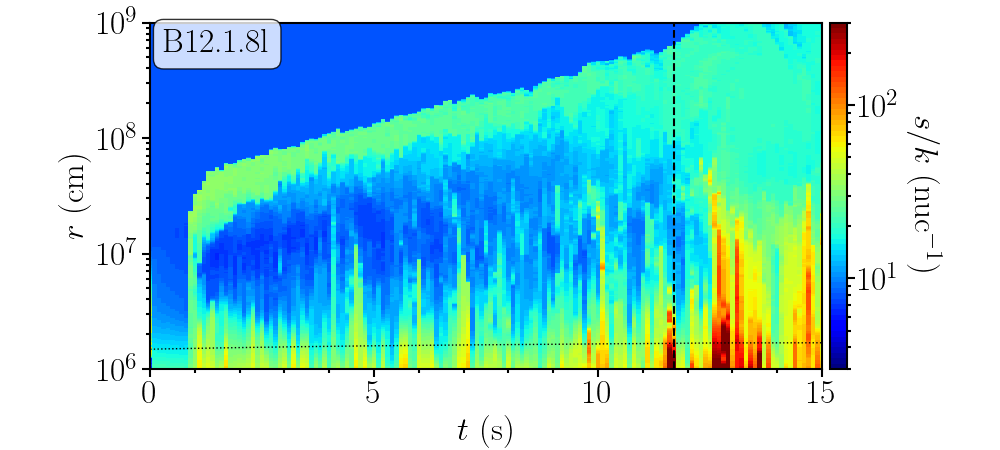}
\includegraphics[width=0.495\textwidth]{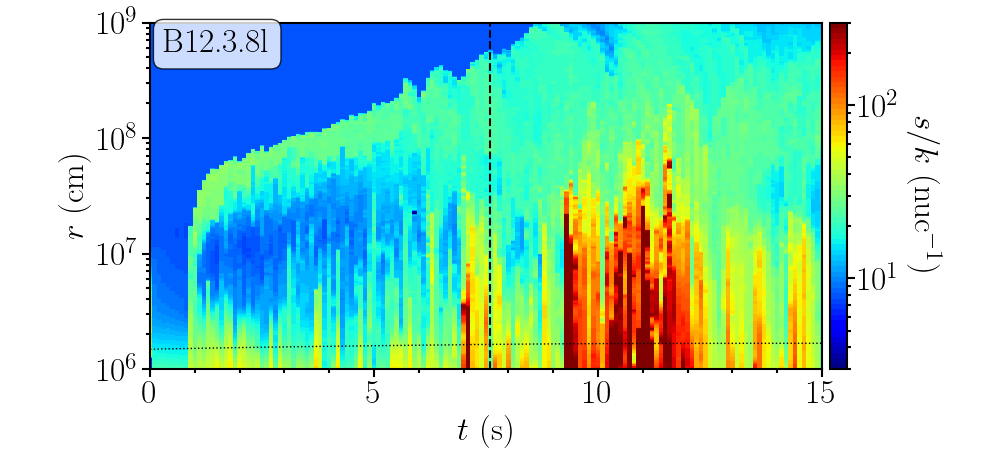}
\includegraphics[width=0.495\textwidth]{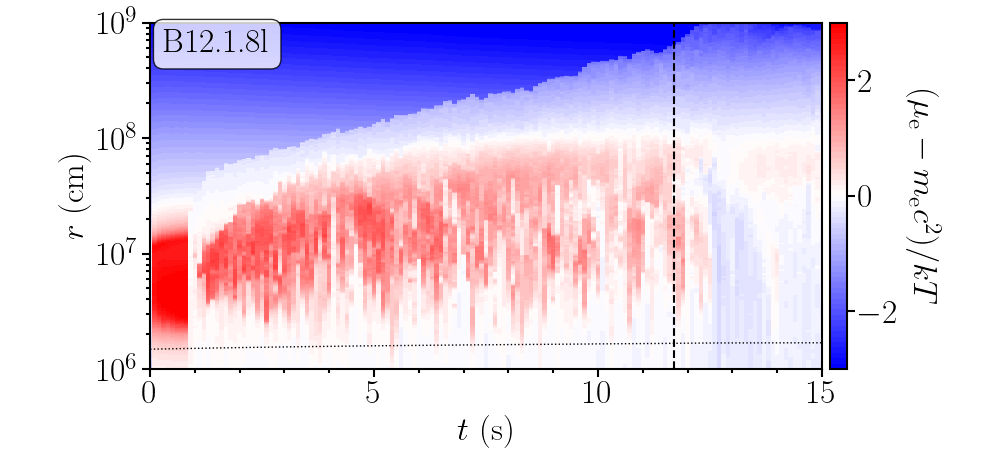}
\includegraphics[width=0.495\textwidth]{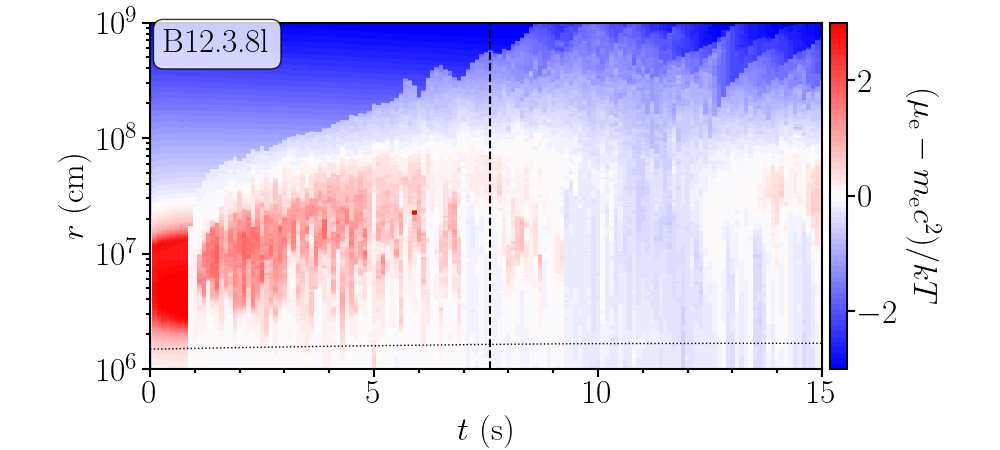}
\includegraphics[width=0.495\textwidth]{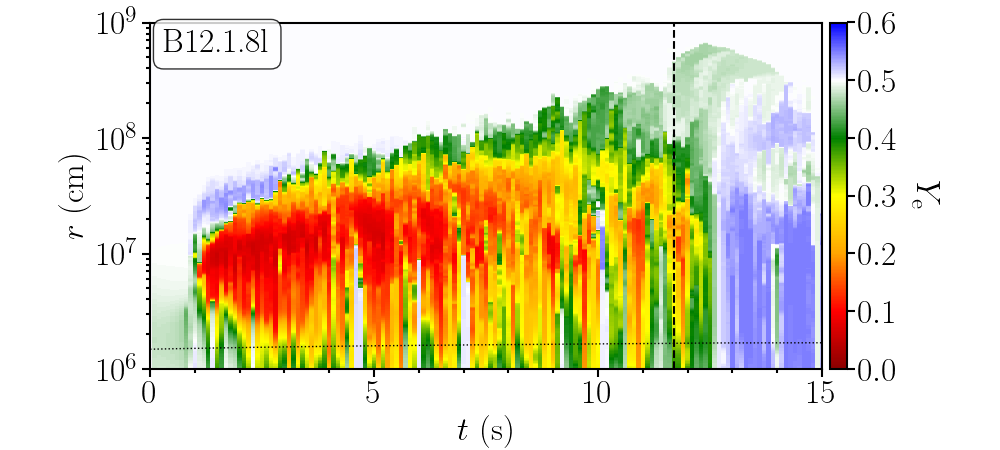}
\includegraphics[width=0.495\textwidth]{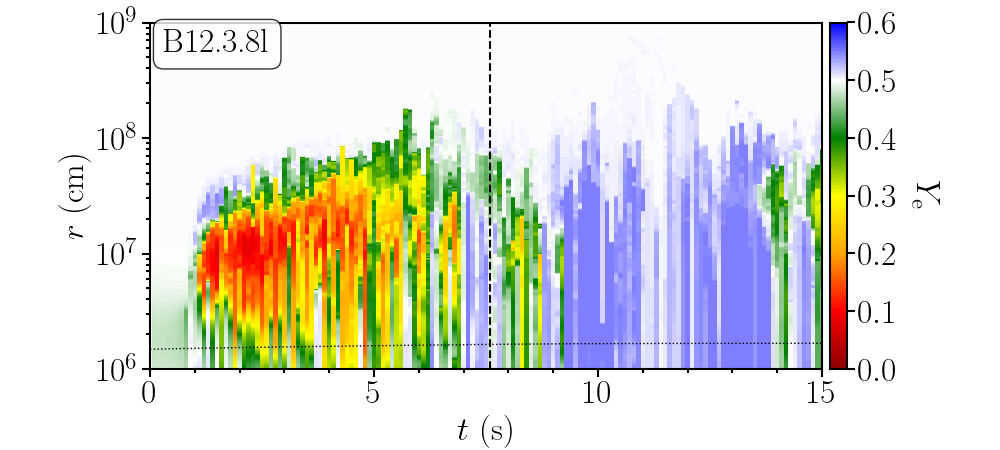}
\caption{Time evolution of various quantities in the equatorial plane. The vertical axis shows the radial coordinate. Top to bottom, the panels show plasma beta, entropy per baryon, electron degeneracy parameter, and electron fraction. The left and right panels correspond to the results for models B12.1.8l and B12.3.8l, respectively. The vertical line in each panel marks the explosion time, and the dotted curve indicates the location of the apparent horizon. $k$ denotes $k_\mathrm{B}$.}
\label{fig:midplane}
\end{figure*}

\begin{figure}
\includegraphics[width=0.495\textwidth]{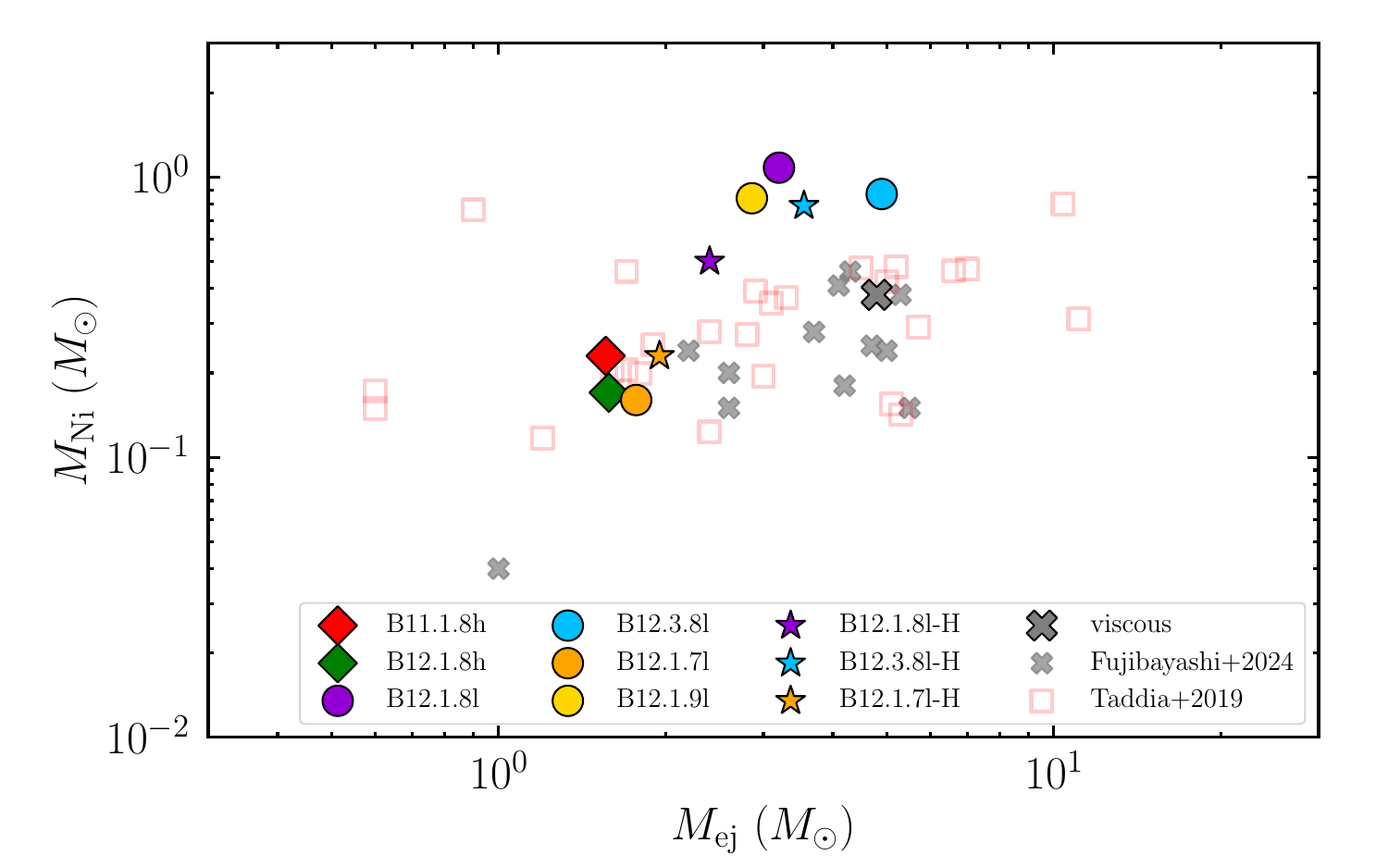}\\
\includegraphics[width=0.495\textwidth]{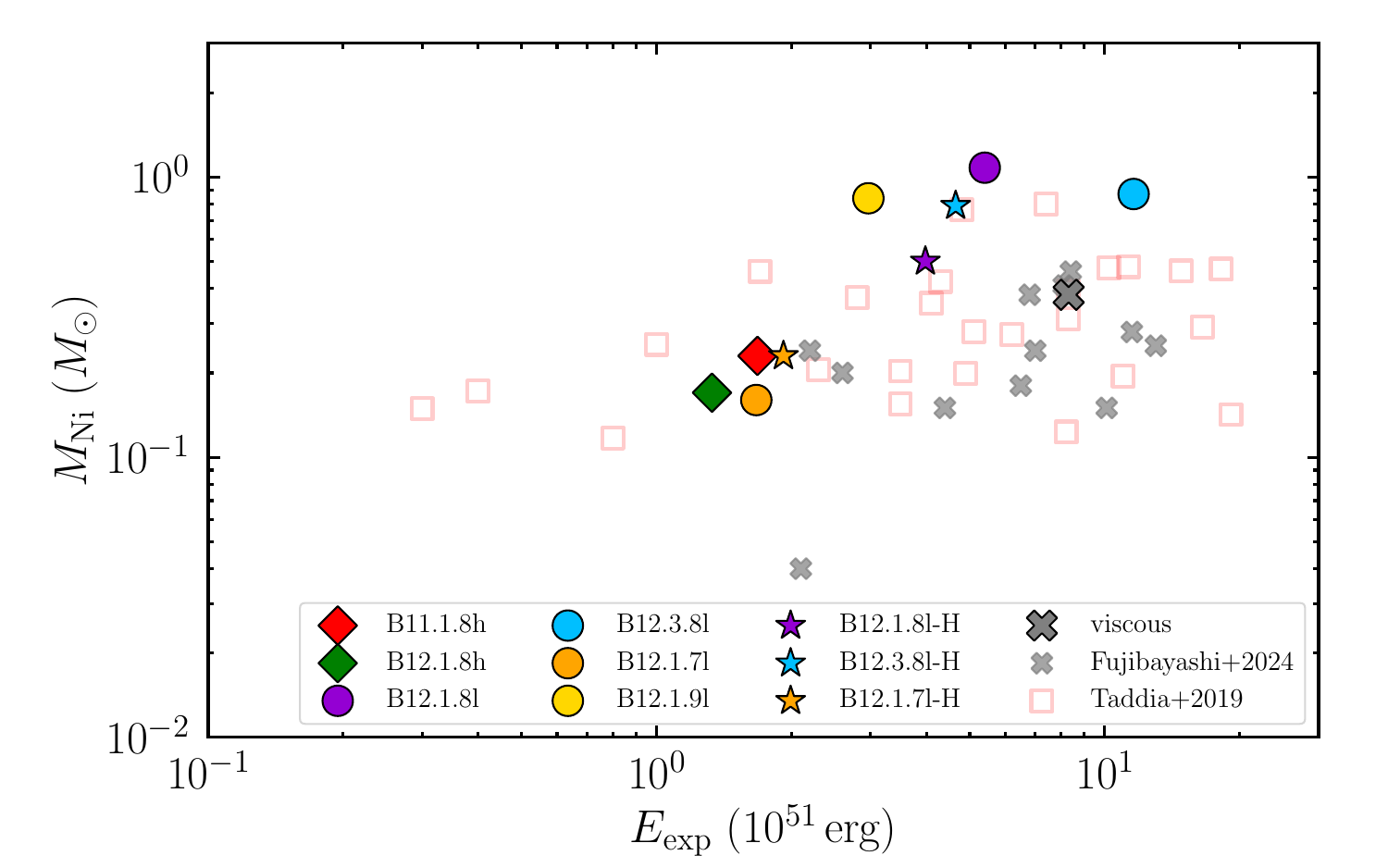}\\
\includegraphics[width=0.495\textwidth]{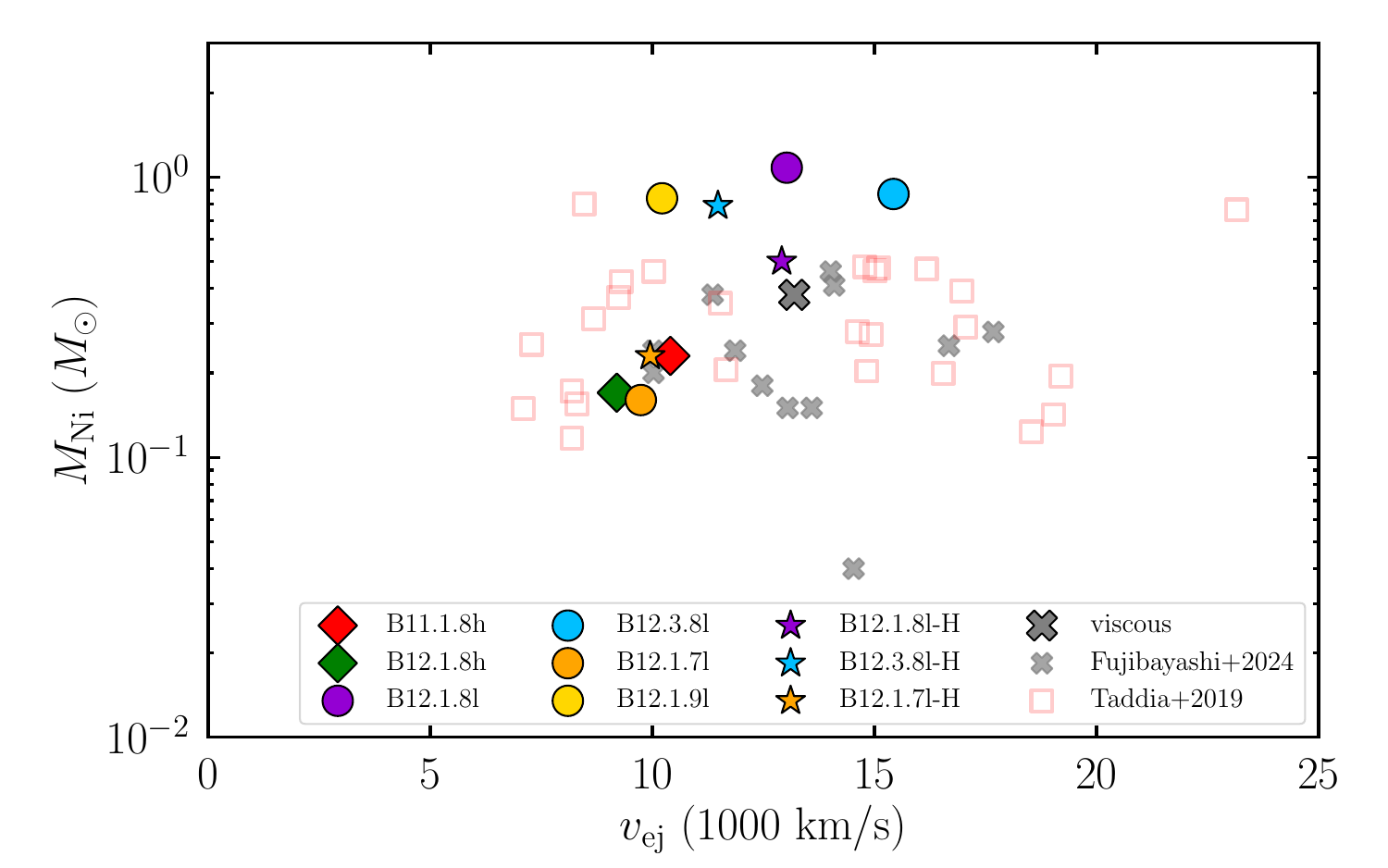}
\caption{Correlations of $^{56}$Ni mass with the ejecta mass, explosion energy, and ejecta velocity. The open markers denote the inferred values from the observations of type Ic-BL SNe \citep{Taddia2019jan}. The grey crosses denote the results of viscous hydrodynamics simulations~\citep{Fujibayashi2024jan}. The diamonds and circles denote the models with high and low cut-off density $\rho_\mathrm{cut}$, respectively. The stars denote the results of the runs with a higher grid resolution.
}
\label{fig:ejecta-summary}
\end{figure}

For nucleosynthesis calculations, we use a tracer particle method to obtain the thermodynamical histories of ejecta. The details of the method were described in Refs.~\cite{Fujibayashi2020c,Fujibayashi:2023oyt}. An order of $10^4$ particles is generated for each calculation.

\subsubsection{Conditions for nucleosynthesis}
Before moving onto the nucleosynthesis results, we first review the general trends of physical conditions found in each of our models. Figure~\ref{fig:cumulative} shows the distributions of the electron fraction (top) and the entropy per baryon (middle) in the ejecta components which experienced the temperature with $>5$ GK (here GK$=10^9$\,K), and the cumulative distributions of the maximum temperature, $T_\mathrm{max}$, that the ejecta ever experienced (bottom). Broadly speaking, for high cutoff density ($\rho_\mathrm{cut}=10^8\,\mathrm{g/cm^3}$; models B11.1.8h and B12.1.8h), the $Y_\mathrm{e}$ distribution is extended to a low value of $Y_\mathrm{e} \alt 0.3$, while for $\rho_\mathrm{cut}=10^6\,\mathrm{g/cm^3}$ (the other models with the letter ``l" in their names), the lowest value of $Y_\mathrm{e}$ is at smallest $\sim 0.3$. The reason for this is that, with the high value of $\rho_\mathrm{cut}$, the turbulence is not excited in the relatively low-density region of the torus with $\rho \alt 10^8\,\mathrm{g/cm^3}$, and hence, the torus with $\rho \alt 10^8\,\mathrm{g/cm^3}$ survives over a long timescale. Because the components with such density have low values of $Y_\mathrm{e}$ due to the presence of degenerate electrons, the more low-$Y_\mathrm{e}$ components can be ejected from the torus for the models with the higher value of $\rho_\mathrm{cut}$. Thus, the appearance of the low-$Y_\mathrm{e}$ components for the models with $\rho_\mathrm{cut}=10^8\,\mathrm{g/cm^3}$ might be an artifact due to our parameter setting, and hence, in the following, we focus mainly on the analysis for the models with $\rho_\mathrm{cut}=10^6\,\mathrm{g/cm^3}$.

The top panels of Fig.~\ref{fig:cumulative} also show that the $Y_\mathrm{e}$ distribution depends on the dynamo parameter $\alpha_\mathrm{d}$ (for a given value of $\sigma_\mathrm{c}$); for $\alpha_\mathrm{d}=3 \times 10^{-4}$ (models B12.3.8l and B12.3.8l-H), the $Y_\mathrm{e}$ values of the ejecta are narrowly distributed between $\sim 0.45$ and $\sim 0.55$, while for $\alpha_\mathrm{d} = 1\times10^{-4}$ (models B12.1.8l and B12.1.8l-H) the range becomes wider ($\sim 0.35$--$0.55$ for the standard resolution-run and $\sim 0.30$--$0.55$ for the higher resolution run). Our interpretation for this is that for the high value of $\alpha_\mathrm{d}$, the dynamo is enhanced more efficiently in the torus, forming a more spread and higher temperature torus. We note that a similar systematic difference was also found in the comparison between models B12.1.7l-H and B12.3.7l-H for which we do not present the results in this paper. 

The top panels of Fig.~\ref{fig:midplane} compare the time evolution of the plasma beta, $8\pi P/B^2$, in the equatorial plane for two models (B12.3.8l and B12.1.8l). The region with small plasma beta $\alt 10$ indeed tends to spread to larger radii at a given time for model B12.3.8l than that for B12.1.8l. At the explosion and jet formation, the density at the inner region of the torus, which subsequently becomes ejecta, is lower while the temperature is higher for model B12.3.8l. As a result, the electron degeneracy is weaker, resulting in the ejection of the matter with higher values of $Y_\mathrm{e}$. This behavior is clearly seen in the second-, third-, and last-row panels of Fig.~\ref{fig:midplane}, in which we compare the mid-plane entropy per baryon, electron degeneracy parameter $(\mu_\mathrm{e}-m_\mathrm{e}c^2)/kT$, and electron fraction, respectively ($\mu_\mathrm{e}$ and $m_\mathrm{e}$ are chemical potential and mass of electron). We note that this trend is seen irrespective of the grid resolution (see, e.g., the top right panel of Fig.~\ref{fig:cumulative}). The results shown here indicate that the $Y_\mathrm{e}$ distribution of the ejected matter depends strongly on the dynamo activity. Specifically, to get low $Y_\mathrm{e}$ ejecta components, a compact torus with strong magnetic fields that ejects the high-density matter in the torus is necessary. 

As the present numerical simulations illustrate, the $Y_\mathrm{e}$ distribution for $Y_\mathrm{e}\alt 0.3$ depends strongly on the details of the dynamo activity. This shows that to obtain more quantitative information on the $Y_\mathrm{e}$ distribution, we need to perform a self-consistent three-dimensional magnetohydrodynamics simulation. 

For $\alpha_\mathrm{d}=1\times10^{-4}$, a relatively low $Y_\mathrm{e}$ component is ejected irrespective of the values of $\sigma_\mathrm{c}$. This component comes from a region of the torus with a relatively high density, for which $Y_\mathrm{e}$ is appreciably lower than $0.5$, by the electromagnetic force. The mechanism is essentially the same as that pointed out in a recent paper~\cite{Issa:2024sts}, but we do not find very low $Y_\mathrm{e}$ components in our simulation with $\rho_\mathrm{cut}=10^6\,\mathrm{g/cm^3}$. This is likely because the poloidal magnetic-field strength is not as high as that in Ref.~\cite{Issa:2024sts}, in which a strong poloidal field is initially given and the mass ejection is artificially enhanced in an early stage of the torus evolution. 

The middle panels of Fig.~\ref{fig:cumulative} shows that a fraction of the ejecta experiences a high-entropy-per baryon state with $\agt 100k_\mathrm{B}$. This is a key to synthesizing a certain amount of $r$-process elements (see the next subsection). We note that for the weak-jet power models, this component is minor and the $r$-process nucleosynthesis becomes weaker. The bottom panels of Fig.~\ref{fig:cumulative} shows that an appreciable fraction of the ejecta experiences a high-temperature state with $T_\mathrm{max} \geq 5$\,GK irrespective of the models. This indicates that a large fraction of $^{56}$Ni can be produced in the ejecta~\cite{Woosley2002a} (see the next subsection). 

The right panels of Fig.~\ref{fig:cumulative} also indicate that a fair convergence of the numerical results with respect to the grid resolution is achieved. One exception is for model B12.1.8l. For the high-resolution run of this model (B12.1.8l-H), an appreciable fraction of the low-$Y_\mathrm{e}$ with $Y_\mathrm{e} \leq 0.35$ is ejected, while for B12.1.8l such low-$Y_\mathrm{e}$ components are absent. The low-$Y_\mathrm{e}$ components come from a deep inside of the torus in a relatively early stage of its evolution at which the density is high (and thus the electron degeneracy is high). For B12.1.8l-H, a strong poloidal magnetic field seems to accidentally penetrate the deep inside of the torus, leading to the ejection of the low-$Y_\mathrm{e}$ component. We have to keep in mind that the convergence of the numerical results might be poor for other models as well. The poor convergence is associated with the fact that the magnetic field strength and structure are determined by the stochastic turbulence state in this problem.

\subsubsection{Nucleosynthesis results}
Based on the thermodynamical histories obtained by the particle tracing method, nucleosynthesis calculations are performed in post-processing. We use a nuclear reaction network code \texttt{rNET}~\cite{Wanajo2018nov}. The network consists of about 6300 nuclear species with atomic number $Z=0$--110, which are connected by relevant reactions. Each nucleosynthesis calculation starts when the temperature decreases to \SI{10}{GK}. The initial abundance is set $1-Y_\mathrm{e}$ and $Y_\mathrm{e}$ for free neutrons and protons, respectively. Choosing such a simple initial composition is justified because the nuclear composition settles into that in nuclear statistical equilibrium (NSE) quickly due to the initial high temperature. If the temperature of a tracer particle never reaches \SI{10}{GK}, the nucleosynthesis calculation starts at the initial time of the simulation. In such a case, the initial composition is set to be that of the pre-collapse star at the radius to which the particle is traced back.

\begin{table*}[t]
\centering
\caption{Mass of ejecta components that experienced above 5\,GK and the masses of $^{44}$Ti, $^{56}$Ni, $^{57}$Ni, Zn, Sr, Te, and lanthanides plus actinides in units of $M_\odot$. The values in the brackets (4th and last columns) show the ratio of the $^{56}$Ni mass with respect to the mass of the matter which experienced $T_\mathrm{max}>5$\,GK and the ratio of the mass of lanthanides plus actinides with respect to the total ejecta mass. 
\label{tab3}
}
\begin{tabular}{lccccccccccc}
\hline\hline
Model & ~~~$>5\mathrm{GK}$~~~ & ~~~$^\mathrm{44}$Ti~~~ & ~~~$^\mathrm{56}$Ni~~~ & ~~~$^\mathrm{57}$Ni~~~ & ~~~Zn~~~ & ~~~Sr~~~ & ~~~Te~~~ & lan.+act.\\
\hline
B12.1.8h   & 0.32  & $2.9\times 10^{-5}$ & ~0.17  (0.51)~ & $2.1\times 10^{-3}$   & $1.2\times 10^{-2}$ & $4.3\times 10^{-3}$  & $1.2\times 10^{-3}$     & ~$1.6\times 10^{-4}$ ($8.6\times 10^{-5}$)~      \\ 
B12.1.8l   & 1.60   & $1.8\times 10^{-4}$ & 1.08  (0.68) & $2.3\times 10^{-2}$   & $8.3\times 10^{-3}$ & $3.1\times 10^{-3}$  & $2.4\times 10^{-5}$     & $5.2\times 10^{-6}$ ($ 1.5\times 10^{-6}$)    \\ 
B12.3.8l   & 1.51  & $1.9\times 10^{-4}$ & 0.87  (0.57) & $7.2\times 10^{-3}$   & $1.9\times 10^{-3}$ & $1.5\times 10^{-4}$  & $9.7\times 10^{-8}$     & $5.2\times 10^{-7}$ ($1.0\times 10^{-7}$)     \\ 
B12.1.7l   & 0.36  & $2.9\times 10^{-5}$ & 0.16  (0.46) & $4.3\times 10^{-3}$   & $2.2\times 10^{-2}$ & $3.6\times 10^{-3}$  &  $5.7\times 10^{-5}$    & $2.2\times 10^{-5}$ ($1.0\times 10^{-5}$)     \\ 
B12.1.9l   & 1.19  & $8.0\times 10^{-5}$ & 0.84  (0.70) & $2.9\times 10^{-2}$   & $9.1\times 10^{-3}$ & $3.6\times 10^{-3}$  &  $1.1\times 10^{-4}$    & $1.0\times 10^{-5}$ ($1.4\times 10^{-5}$)     \\ 
\hline
B12.1.8l-H   & 1.09  & $9.6\times 10^{-5}$ & 0.50 (0.46) & $1.0\times 10^{-2}$   & $3.6\times 10^{-2}$ & $1.4\times 10^{-2}$  & $1.2\times 10^{-3}$   & $5.1\times 10^{-4}$ ($1.8\times 10^{-4}$)     \\ 
B12.3.8l-H   & 1.22  & $1.1\times 10^{-4}$ & 0.79  (0.65) & $2.1\times 10^{-2}$   & $1.0\times 10^{-3}$ & $7.0\times 10^{-9}$  &  0  & 0      \\ 
B12.1.7l-H   & 0.54  & $5.1\times 10^{-5}$ & 0.23  (0.43) & $4.3\times 10^{-3}$   & $4.2\times 10^{-2}$ & $2.6 \times 10^{-3}$  & 
$2.2\times 10^{-4}$  & $6.7\times 10^{-5}$ ($2.8\times 10^{-5}$)     \\ 
\hline
viscous & 1.00 & $5.5\times 10^{-5}$ & 0.42 (0.42) & $1.6\times 10^{-2}$   & $5.8\times 10^{-4}$ & $1.8\times 10^{-10}$ & 0 &  0     \\ \hline
\end{tabular}\\
\end{table*}

Table~\ref{tab3} lists several important quantities of the nucleosynthesis results. The mass of the synthesized $^{56}$Ni amounts to 0.1--1.1$M_\odot$, reflecting the mass of the matter for which $T_\mathrm{max} \geq 5$\,GK. The ratio $M_\mathrm{Ni}/M_{>5\,{GK}}$ spans from $\approx 0.35$ to $\approx 0.7$. This difference in the production efficiency of $^{56}$Ni stems from the difference in the entropy and electron fraction of the ejecta; the conditions with low entropy and high electron fraction ($Y_\mathrm{e} \gtrsim 0.49$) are favored for the efficient $^{56}$Ni production. 

Broadly speaking, the $^{56}$Ni mass is larger for more energetic explosion models with larger values of $M_\mathrm{ej}$ and $E_\mathrm{exp}$. 
Figure~\ref{fig:ejecta-summary} plots correlations between the ejecta mass and the $^{56}$Ni mass (top), between the explosion energy and the $^{56}$Ni mass (middle), and between the ejecta velocity and the $^{56}$Ni mass (bottom). Here, the ejecta velocity is defined by
\begin{align}
\frac{v_\mathrm{ej}}{c} := \sqrt{1-\Gamma^{-2}},
\end{align}
where $\Gamma := E_\mathrm{exp}/M_\mathrm{ej}$ is the terminal Lorentz factor of the ejecta, and we find $10^4\,\mathrm{km/s} \alt v_\mathrm{ej} \alt 1.5\times 10^4\,\mathrm{km/s}$ in our present models. Together with the numerical data, the inferred values from the observations of type Ic-BL SNe \citep{Taddia2019jan} as well as the result of viscous hydrodynamics simulations~\citep{Fujibayashi2024jan} are also plotted. As in Fig.~\ref{fig:ejecta-Eene}, it is found that our results are in a good agreement with the observational results for $M_\mathrm{ej}\approx 1$--$5M_\odot$, $E_\mathrm{exp} \approx 10^{51}$--$10^{52}$\,erg, and $v_\mathrm{ej} \approx 10^4$--$1.5\times 10^4$\,km/s, indicating that our explosion scenario (explosion launched from a massive torus around a spinning black hole) is a robust model for interpreting the energetic SNe. 


Given that our models represent at least in part the observed type Ic-BL SNe, radioactive isotopes other than $^{56}$Ni, such as $^{44}$Ti and $^{57}$Ni, are expected to play a role in the late-time light curves as in the case of SN~1987A (from $\sim 2000$ and $\sim 1000$~days for the former and the latter, respectively \cite{Seitenzahl:2014zia}). Our models predict the $^{44}$Ti mass between $3\times 10^{-5} M_\odot$ and $2\times 10^{-4} M_\odot$, which is consistent with the inferred amounts for SN~1987A (($3.1\pm 0.8)\times 10^{-4} M_\odot$ \cite{Grebenev:2012ck}) and Cas~A (($1.25\pm 0.3)\times 10^{-4} M_\odot$ \cite{Grefenstette:2014fga}) by gamma-ray spectroscopy. Our models also predict the ratio of $^{57}$Ni/$^{56}$Ni $\sim 0.01$--0.04, or [$^{57}$Ni/$^{56}$Ni] $\equiv [M(^{57}\mathrm{Ni})/M(^{56}\mathrm{Ni})]/[M(^{57}\mathrm{Fe})/M(^{56}\mathrm{Fe})]_\odot \sim 0.6$--1.7 ($M(Q)$ is the mass of the indicated isotope $Q$), which is consistent with the inferred value for SN~1987A ([$^{57}$Ni/$^{56}$Ni] $= 1.5 \pm 0.3^\mathrm{stat} \pm 0.2^\mathrm{sys}$ \cite{Kurfess1992}). To confirm the presence of these radioactive isotopes in the late-time light curves of type Ic-BL SNe, a long duration of observations over a few 1000 days will be needed (e.g., the data only up to $\sim 500$~days exist for SN~1998bw \cite{Clocchiatti:2011wa}).

\begin{figure*}[t]
\includegraphics[width=0.49\textwidth]{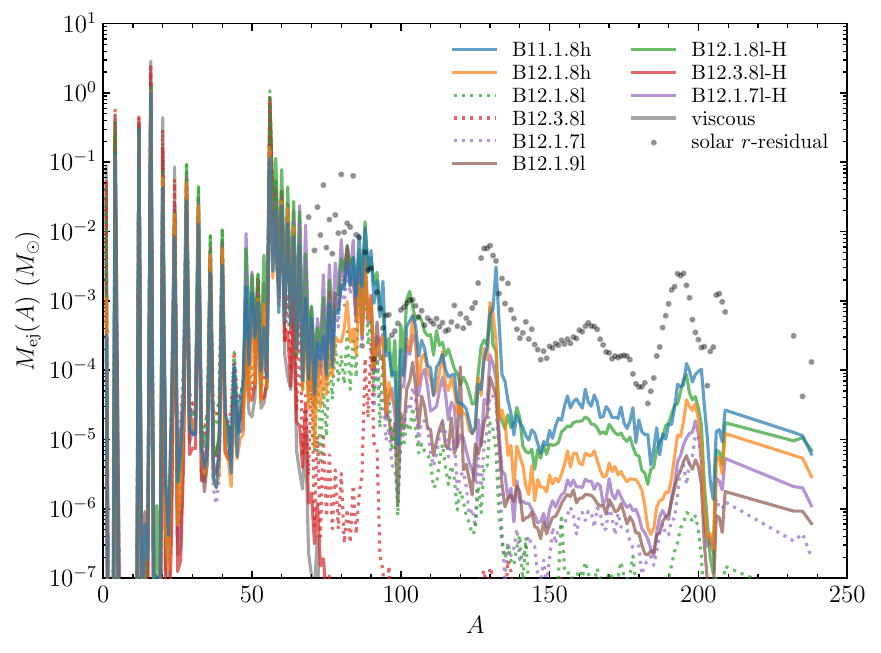}
\includegraphics[width=0.49\textwidth]{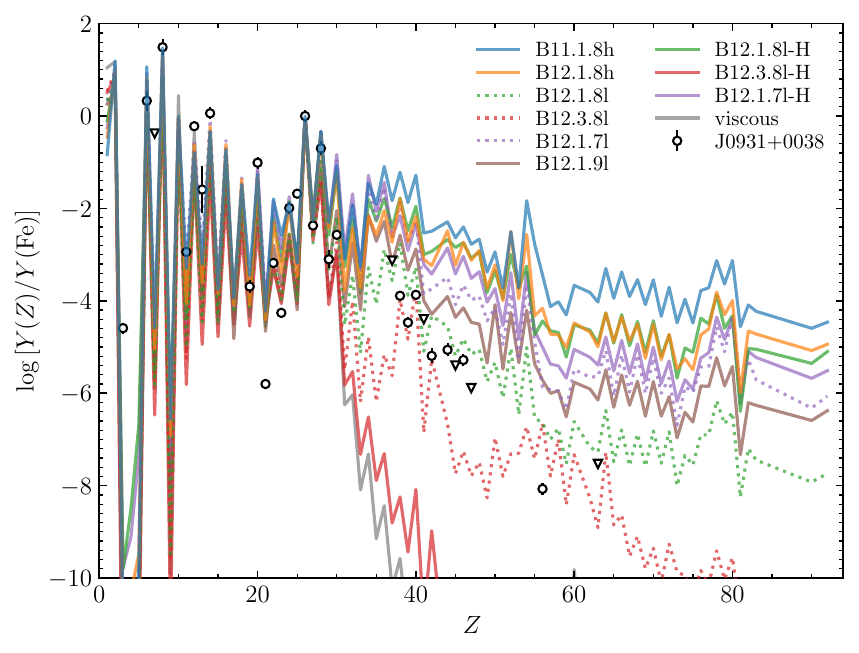}
\caption{Left: Ejected masses of nuclei in units of $M_\odot$ for all models as a function of atomic mass number $A$ shown by the solid curves with different colors indicated in the legend. The dotted curves show the results for the standard-resolution runs corresponding to the high-resolution runs indicated by ``-H". The circles show the $r$-process residuals to the solar abundances \citep{Prantzos2020a}, which are vertically shifted to match the result for B11.1.8h at $A = 90$. Right: Elemental abundances for all models normalized by those of Fe. The circles with error bars are the measured stellar abundances of J0931+0038, which is suggested to have been polluted by a single explosive event with the initial progenitor mass $> 50 M_\odot$ \citep{Ji:2024edf}.
}
\label{fig:abundance}
\end{figure*}

Figure~\ref{fig:abundance} shows the abundance distributions obtained from our nucleosynthesis calculations. The left panel displays the ejecta masses of synthesized nuclei as functions of atomic mass number $A$ for all models, which are compared to the $r$-process residuals to the solar abundance \cite{Prantzos2020a} (scaled to match the result for B11.1.8h at $A = 90$). Note that the dominance of $\alpha$ elements for $A \le 40$ reflects the adopted initial composition of the pre-collapsing star, not due to the result of explosive nucleosynthesis. We find that the presence of a jet (see the last column in Table~\ref{tab1}) appears to be essential in synthesizing trans-iron species. In fact, no trans-iron nuclei are produced in the non-jet models (B12.3.8l-H and viscous), while some $r$-process nuclei are synthesized in the other models. However, the $r$ process is overall very weak, forming only up to the second peak of $A \sim 130$, which is similar to our previous work for the post-neutron-star-merger remnants~\cite{Fujibayashi2020b} but different from the prediction in Ref.~\cite{Siegel2019may}. Note that a strong $r$ process can occur in the models with jets under moderately low-$Y_\mathrm{e}$ ($\sim 0.3$--0.49) conditions because of the presence of the high-entropy ($> 100 k_\mathrm{B}$ per baryon) components \cite{Hoffman:1996aj,Fujibayashi2020b}. However, because of the small masses of such high-entropy components (the middle panels of Fig.~\ref{fig:cumulative}), the mass-averaged abundances exhibit weak $r$-process-like abundance trends.

The 6th column in Table~\ref{tab3} lists the ejected mass of Zn ($Z = 30$) for all models. The astrophysical origin of Zn is still unknown \cite{Kobayashi:2006nc, Hirai:2018foc}, although hypernovae \cite{Umeda:2001kd,Tominaga:2007pb} and electron-capture SNe \cite{Wanajo2011a,Wanajo2018jan} are suggested to be possible production sites. It is noteworthy that our models predict quite large amounts of Zn (mostly $^{64}$Zn, $\sim 0.001$--$0.04 M_\odot$), in which the lower bound is consistent with those suggested in the previous studies \cite{Umeda:2001kd,Tominaga:2007pb,Wanajo2011a,Wanajo2018jan}. We find the higher Zn mass for the models with jets, in which the high entropy leads to a strong alpha-rich freeze-out, resulting in the appreciable production of $^{64}$Zn. Thus, our result implies that type Ic-BL SNe (or hypernovae) might be (in part) the astrophysical sources of Zn. 

The three last columns in Table~\ref{tab3} list the ejecta mass of Sr ($Z = 38$, as representative of the first $r$-process peak nuclei), Te ($Z = 52$, as representative of the second $r$-process peak nuclei), and the sum of lanthanides and actinides. We find a variation of $r$-process productivity among different models with a tendency of a larger amount of $r$-process nuclei for the high-$\rho_\mathrm{cut}$ models (B11.1.8h and B12.1.8h), although model B12.1.8l-H exhibits a similar productivity to these models. It is interesting to note that the $r$-process nuclei are absent in the viscous model, which is similar to the result of a dynamo model B12.3.8l-H with no jet formation. Thus, our result implies that the jet formation is the key to synthesizing $r$-process nuclei and the emission line of Te might be detected in the light curves of GRB-associated type Ic-BL SNe as in the cases of the two kilonovae with the Te feature \cite{JWST:2023jqa,Hotokezaka:2023aiq}, while such a feature would be absent for those with no GRB. However, because of the small amounts of lanthanides and actinides (typically with the mass fraction below $10^{-4}$; the last column, in the brackets of Table~\ref{tab3}), the type Ic-BL SNe represented by our models would not be observed as kilonova-like red transients, which is consistent with the indication of no or very little amounts of $r$-process elements in the ejecta of GRB-associated type Ic-BL SNe \cite{Rastinejad:2023ipp}. Our present results suggest that the source of the kilonova associated with the long-duration GRB~230307A \cite{JWST:2023jqa} is unlikely a collapsing massive star but probably a neutron-star merger.

\begin{figure*}[t]
\includegraphics[width=0.495\textwidth]{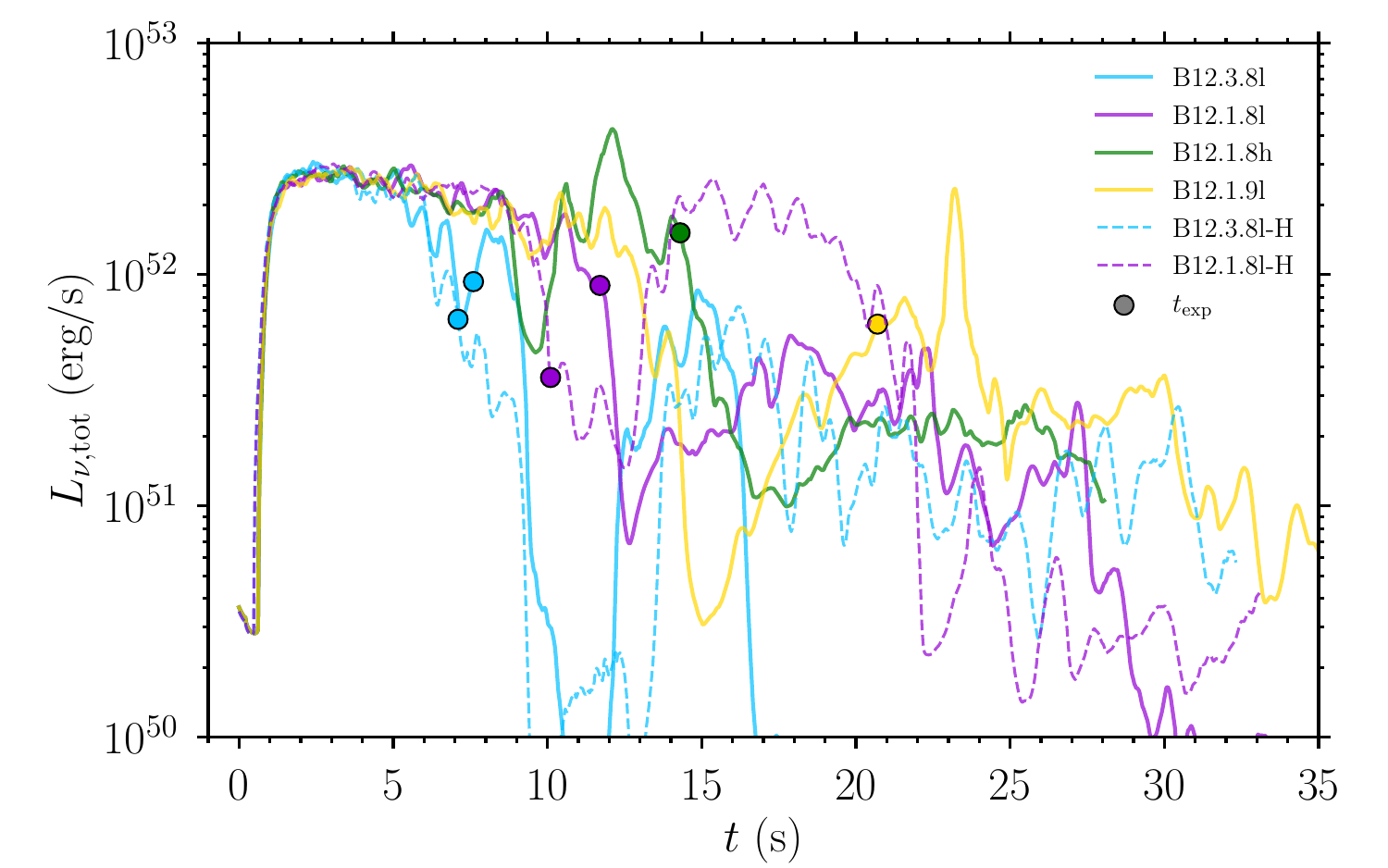}
\includegraphics[width=0.495\textwidth]{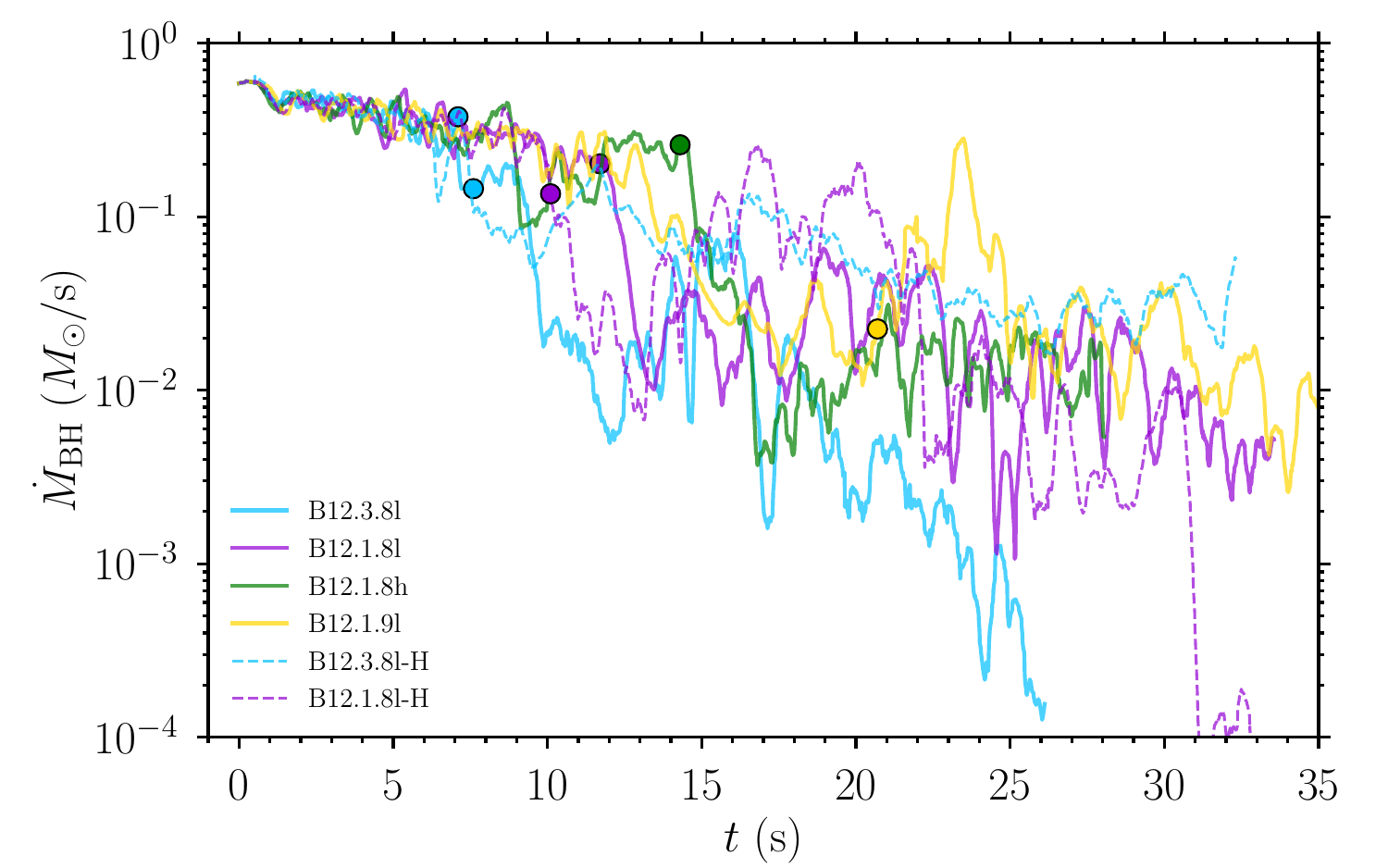}
\caption{Left: Total neutrino luminosity for selected models, in which the explosions set in at earliest (B12.3.8l), latest (B12.1.9l), and middle (B12.1.8l and B12.1.8h), respectively. The results for high-resolution runs are also shown if present. Right: Mass accretion rate onto the black hole for the same models as the left panel. For both panels, the circles indicate the explosion time.}
\label{fig:lnutot}
\end{figure*}

The right panel of Fig.~\ref{fig:abundance} displays the elemental abundance $Y$ (number per baryon) as a function of atomic number $Z$, which is normalized by that of Fe for each model. We find a bifurcation of productivity beyond iron; no $r$ process for the models without jets (B12.3.8l-H and viscous) and a weak $r$ process for the models with jets. The abundance pattern of such a weak $r$ process can be an explanation for the descending trend of trans-iron elements found in Galactic metal-poor stars \cite{Honda2006a}. It is interesting to note that model B12.3.8l exhibits the abundance distribution in between this bifurcation of trans-iron production. In fact, the result for this model is in a reasonable agreement with the very unusual abundance pattern of a recently discovered metal-poor star J0931+0038 \cite{Ji:2024edf}. It is suggested that the unusual abundance trend in this star, namely, an extreme odd-even pattern, the high abundance of Zn, and the small abundances of $r$-process elements, reflects the single nucleosynthetic event of a star with initial mass $> 50 M_\odot$, either of a pair-instability SN or a type Ic-BL SN (hypernova) \cite{Ji:2024edf}. Our result suggests that a type Ic-BL SN with similar conditions to model B12.3.8l might be the source of this unusual abundance pattern. More examples of such rare abundance patterns in metal-poor stars in upcoming spectroscopic surveys will serve to constrain our models.

\subsection{Neutrino emission and memory gravitational waves}

The left panel of Fig.~\ref{fig:lnutot} shows the total neutrino luminosity for selected models. The sudden rise seen at $t\approx \SI{1.5}{s}$ for all these models is due to the hot and massive torus formation. Until $t\approx \SI{5}{s}$, they show similar behaviors of the luminosity. The luminosity begins to fluctuate when the turbulence is significantly excited inside the torus and after the stellar explosion, the neutrino luminosity sharply drops (see the circles). 

A correlation is found between the neutrino luminosity and the mass accretion rate of the black hole (see the right panel of Fig.~\ref{fig:lnutot}). Broadly speaking, before the stellar explosion, the rate of the mass accretion onto the black hole remains high, but after the explosion sets in, it drops. As a consequence, the neutrino luminosity drops as well. For models with earlier explosion such as B12.3.8l and B12.3.8l-H, this feature is clearly seen. For such models the total energy emitted by neutrinos becomes small (see Table~\ref{tab2}). 

The entire evolution of the light curve of the neutrino luminosity from the onset of the core collapse to the stellar explosion should be unique in the present collapsar scenario: It is expected that at the formation of a proto-neutron star, a neutrino burst with the total luminosity of order $10^{52}$\,erg/s should take place, and the high-luminosity stage would continue until the formation of a black hole (e.g., Refs.~\citep{Janka2012a, Sumiyoshi2006sep, Kuroda2023nov}); for the next few tens seconds, the black hole simply grows without the formation of a disk/torus, and hence, the neutrino luminosity is expected to be much lower than $10^{52}$\,erg/s; however, after the formation of a massive torus, a high-luminosity state appears again as shown in Fig.~\ref{fig:lnutot} (see also the middle panel of Fig.~4 of Ref.~\cite{Fujibayashi:2022xsm} for the neutrino light curve). If this unique luminosity curve for neutrinos could be detected in future, it can be an evidence for the collapsar scenario described in this paper. 

After the formation of a massive torus, the neutrino luminosity is enhanced beyond $10^{52}\,\mathrm{erg/s}$. This high neutrino luminosity stage continues for $\Delta t= 10$--20\,s until the onset of the stellar explosion, and thus, the total radiated energy can be beyond $10^{53}$\,erg (see Table~\ref{tab2}). The neutrinos are emitted from the torus of an anisotropic structure and the emission direction should be naturally anisotropic. In the presence of the anisotropic emission of radiation, memory-type gravitational waves are emitted (see Refs.~\cite{1978ApJ...223.1037E, 1978Natur.274..565T, Mueller2012}). An order of the magnitude estimation gives the amplitude of gravitational waves as
\beq
h \approx 4 \times 10^{-24} \left({200\,\mathrm{Mpc} \over D}\right)
\left({E_\nu \over 3 \times 10^{53}\,\mathrm{erg}}\right)
\left({\epsilon_\mathrm{ani} \over 10^{-1}}\right),
\eeq
where $D$ is the distance to the source, $E_\nu$ the total energy emitted by neutrinos, and $\epsilon_\mathrm{ani}$ the degree of the anisotropy of the emission. The typical frequency of the gravitational waves is $1/\Delta t \alt 0.1$\,Hz. The frequency is too low to be the source of ground-based gravitational-wave detectors such as advanced LIGO. However, this can be a source for a future high-sensitivity space detector such as DECIGO~\cite{Kawamura:2020pcg}.


\section{Summary and conclusion}\label{sec4}

We reported the results for the collapsar models, which were obtained by performing neutrino-radiation resistive magnetohydrodynamics simulations with a dynamo term for a massive rotating progenitor star. For many of our models, we found the jet launch and stellar explosion together. Both the jet launch and stellar explosion were induced after the development of a magnetohydrodynamical turbulent state of the torus, for which 10--20\,s evolution after the onset of the disk/torus formation is necessary. A summary for the order of the timescales from the onset of the collapse through the core bounce, the black hole formation, the disk/torus formation, jet launch, and stellar explosion is presented in Table~\ref{tab:summary}. In this scenario, the onset of the stellar explosion takes place at $\agt 10$\,s after the onset of the stellar core collapse in contrast to ordinary SNe, for which the explosion would be induced within 1\,s after the onset of the stellar core collapse. 

\begin{table}[t]
\centering
\caption{Order of the relevant timescales after the onset of the collapse in the collapsar scenario presented in this paper. PNS denotes proto-neutron star. 
\label{tab:summary}
}
\begin{tabular}{lc}
\hline\hline 
 & ~~Order of Timescale (s)~~ \\
\hline
Core bounce/PNS formation   & $O(0.1)$  \\
Black hole formation & $O(1)$ \\
Onset of disk formation & $O(10)=\tau_\mathrm{df}$ \\
Jet \& explosion & $O(10) + \tau_\mathrm{df}$ \\
Reconnection & $\alt 100$\,s\\
\hline
\end{tabular}\\
\end{table}

As we discussed in our previous paper~\cite{Shibata:2023tho}, the extraction of the rotational kinetic energy of a spinning black hole could continue as long as the poloidal magnetic fields that penetrate the black hole is present. Since the rotational kinetic energy of black holes with moderate magnitude of spin is beyond $10^{53}$\,erg, which is much larger than the typically emitted total energy of GRBs including the afterglow and stellar explosion, the extraction process has to be stopped before the entire rotational kinetic energy of the black hole is extracted. This would require the decay of the poloidal magnetic field that penetrates the black hole. Since the typical duration of the long GRBs is at most 100\,s, the decay should occur in a similar timescale (see a discussion in Ref.~\cite{Shibata:2023tho}). We infer that the decay will be achieved by a reconnection process of magnetic field lines, which is also listed in Table~\ref{tab:summary}. 

The detailed mechanisms of the jet launch and stellar explosion found in this paper are summarized as follows: After the formation of a massive torus, a turbulent state is developed by the dynamo action, and as a result, the magnetic-field strength is enhanced in it. Together with the angular momentum transport associated with the effective viscosity resulting from the turbulence and the growth of the torus due to the matter infall from the outer envelope, the torus expands due to the (effective) viscous heating with time. However, in an early stage, the explosion by the heating in the torus is suppressed by the ram pressure of the infalling matter. Matter accretes from the torus to the black hole, and associated with this, the magnetic flux is provided to the black hole. However, in the early stage of the evolution, the ram pressure of the matter infalling onto the black hole is so strong that the magnetic field is swallowed into the black hole without developing a magnetosphere that is suitable for launching an energetic jet via the Blandford-Znajek effect. 

These situations change in a later stage in which the density of the infalling matter decreases. As a result of this, its ram pressure decreases and the torus expansion is enhanced. In particular the magnetohydrodynamics effect and (effectively) viscous heating in the region close to the black hole plays a central role in the energy injection. Also, developing a geometrically thick torus, which can block the matter infall to the central region near the black hole, becomes possible. When the electromagnetic force (such as magnetic pressure) near the black hole, enhanced by the magnetic winding associated with the black hole spin, overcomes the ram pressure, a jet is eventually launched, leading to the development of a magnetosphere. Also, the stellar explosion associated primarily with the viscous heating in the torus is induced after the ram pressure from the infalling matter onto the torus becomes sufficiently low. The mechanism for the stellar explosion is qualitatively identical with that in viscous hydrodynamics~\cite{Fujibayashi:2023oyt}. 

For the present models, the explosion energy of the star is $10^{51}$--$10^{52}$\,ergs depending on the parameters of the dynamo term. For the models in which a jet is launched earlier (in $\sim 10$\,s after the formation of a disk), the explosion energy is much higher than $10^{51}$\,erg and can explain luminous SNe such as type Ic-BL SNe (or hypernovae)~\cite{Cano2017a}. For such models, the synthesized $^{56}$Ni mass far exceeds $0.1M_\odot$, suitable for the model of type Ic-BL SNe. Even for late-time explosion models for which the explosion energy is relatively low, the $^{56}$Ni mass becomes $0.1$--$0.2M_\odot$, which can account for bright SNe. We also found a large amount of Zn production ($\sim 0.001$--$0.04M_\odot$). Our result suggests the astrophysical origin for Zn being such energetic SNe.

We found that only in the presence of a jet, $r$-process elements could be synthesized, because low-$Y_\mathrm{e}$ and high-entropy matter is ejected from a dense region of the torus by the magnetohydrodynamical effect. However, in our present models, the $r$-process is so weak that the elements are likely to be sufficiently synthesized only up to the second peak; only a small amount of lanthanoids and actinides are likely to be synthesized. This suggests that the red-kilonova-type observation would not be expected in our collapsar model. Thus, our models cannot explain a recently discovered kilonova-like transient associated with the long-duration GRB~230307A \cite{JWST:2023jqa}. Such weak $r$-process-like abundance trends with large amounts of Fe and Zn can be explanations for some peculiar abundance patterns found in metal-poor stars \cite{Ji:2024edf}.

Recent analysis for the observational results of SNe~\cite{2023ApJ...953..179C} shows that a large fraction of type Ic SNe does not accompany the relativistic motion such as jets. Thus, our model presented in the present paper does not represent the general mechanism for the type Ic-BL SNe. However, the energy injection from a massive torus is  still a viable mechanism. Thus, some of type Ic SNe may be powered by the massive torus surrounding a black hole, with no jet launching.

Because of the simplification for the modeling of the dynamo effect, the development of the magnetosphere after the jet launch might not be well modeled because the magnetic-field strength decays with time of order 10\,s in the polar region. As a result, the Poynting luminosity might be underestimated in the current model. Nevertheless, the total energy of the Poynting luminosity can be $10^{50}$--$10^{51}$\,ergs for all the models, which are suitable values for explaining long GRBs~\cite{2008ApJ...675..528L}. In any case, a simulation with more sophisticated modeling for the dynamo effect or a three-dimensional ideal magnetohydrodynamics simulation in which the dynamo effect is captured is necessary for a more quantitative understanding of the collapsar scenario in the future.

\acknowledgements

We thank David Aguilera-Dena for providing their stellar evolution models. We also thank Alessandra Corsi, Kenta Hotokezaka, and Kohta Murase for useful discussions. Numerical computation was performed on Sakura and Momiji clusters at Max Planck Computing and Data Facility. This work was in part supported by Grant-in-Aid for Scientific Research (grant Nos.~20H00158 and 23H04900) of Japanese MEXT/JSPS.

\bibliography{reference}

\end{document}